\documentclass[10pt,a4paper]{article}
\usepackage{amsmath}
\usepackage{amsthm}
\usepackage{epsfig}
\usepackage{latexsym}
\usepackage{xcolor}
\usepackage{pdflscape}
\numberwithin{equation}{section}
\usepackage{graphicx}
\usepackage{dcolumn}

\title{\Large \bf Investigating the configurations in cross-shareholding:\\  a joint copula-entropy approach}

\author{Roy Cerqueti $^{1,\ddagger}$, Giulia Rotundo $^{2, \dagger,\ddagger}$* and Marcel Ausloos $^{3,\ddagger}$}

\date{
$^{1}$ \quad Department of Economics and Law, University of Macerata, via Crescimbeni, 20, 62100 Macerata, IT
\\email address: roy.cerqueti@unimc.it\\
$^{2}$ \quad Department of Statistical Sciences, Sapienza University of Rome, p.le A. Moro 5, 00185 Roma, IT
 \\email address: giulia.rotundo@uniroma1.it\\
$^{3}$ \quad School of Business, University of Leicester,
University Road, Leicester, LE1 7RH, UK
 \\email address: ma683@le.ac.uk}
\begin{document}
 \maketitle


\abstract{The complex nature of the interlacement of economic actors
is quite evident at the level of the Stock market, where any company
  may actually interact  with the other companies
buying   and selling their shares. In this respect, the companies
populating a Stock market, along with their connections, can be
effectively modeled through a directed network, where the nodes
represent the companies, and the links indicate the ownership. This
paper deals with this theme and discusses the concentration of a
market. A cross-shareholding matrix is considered, along with two
key factors: the node out-degree distribution which represents the
diversification of investments in terms of the number of
involved companies, and the node in-degree distribution which
reports the integration of a company due to the sales of its own
shares to other companies. While diversification is widely explored
in the literature, integration is most present in literature on
contagions. This paper captures such quantities of interest in the
two frameworks and studies the stochastic dependence of
diversification and integration through a copula approach. We adopt
entropies as measures for assessing the concentration in the market.
The main question is to assess the dependence structure
leading to a better description of the data or to market
polarization (minimal entropy) or market fairness (maximal
entropy). In so doing, we derive information on the way in which
the in- and out-degrees should be connected in order to shape the
market. The question is of interest to regulators bodies, as
witnessed by specific alert threshold published on the US mergers
guidelines for limiting the possibility of acquisitions and the
prevalence of a single company on the market. Indeed, all countries
and the EU have also rules or guidelines in order to limit
concentrations, in a country or across borders, respectively. The
calibration of copulas and model parameters on the basis of real
data serves as an illustrative application of the theoretical
proposal.}



\section{Introduction}
The recent crises have evidenced the fragility of the financial
system due to the growing interdependencies among many different
organizations.

In the context of network modeling applied to management
organizations of industrial structures, usually nodes represent
companies, while the links show the ownership, gathered in the
cross-shareholding matrix. However, many studies in literature
mostly focused on the shape of the distribution of the node
out-degree $k_{out}$, because such results are linked to specific
results on the resilience of the network \cite{Newman, soromaki,
gao, iori, delpini}. $k_{out}$ represents the number of the
companies whose stocks are included in the portfolio of the
considered company, i.e. it is the amount of different counterparts.
Therefore, $k_{out}$ can be used for representing the
diversification, according to its conceptualization in the reference
literature (see e.g. \cite{SoumaBook}). The higher  the
diversification, the less sensitive the node is to its inner
fluctuations.

Surprisingly, not many studies were done on the node in-degree
$k_{in}$ distributions, where $k_{in}$ is the amount of (other)
companies who bought some ownership of a specific company. The
in-degree well represents the way in which each organization becomes
more dependent on its counterparts, so it can be used to represent
the integration of the company in the system (also for the
concept of integration, refer to \cite{SoumaBook}).

Notice that the construction of $k_{out}$ and that of
$k_{in}$ do not involve the entity of the connections among
companies, but only the number of existing connections. Thus, such
quantities serve for modeling the presence of interactions;  this
provides information on how a company is integrated in the system
and how diversified is its portfolio.

An initial increase of integration may allow financial fluctuations
of the value of a company to propagate  and very high integration allows
eventual cascades to spread on so many units that its effects are
minimal \cite{Elliott}. Literature contributions inquired furtherly on  on the trade-off among integration and
differentiation so to detect the most dangerous combination for the
propagation of a global crisis \cite{Elliott}. In this respect, it is also worth
mentioning other ways  for interconnections among companies, like
the interlock of directorates \cite{Bellenzier,Croci,GRAMDAASS} or
personal relationships \cite{Gulati}, or other contractual
relationship (for a survey, see \cite{weber}).

However, it is important to stress once again that $k_{out}$ is much
more studied than $k_{in}$ in the empirical literature (see the
review below).

Studies on different real world networks have shown different
reactions to patterns of attack among highly versus low concentrated
networks. In short, highly concentrated networks are resilient to
random shocks, but most sensitive to attacks to the core and to
hubs. On the opposite, low concentrated networks are sensitive to
random
attacks \cite{Pastor,Pastor1}. 

In this paper, we elaborate on the market concentration, represented
through the entropies of the distributions of diversification and
integration. In a connected network, under the hypothesis of
independence among $k_{in}$ and $k_{out}$, the entropy is minimal
when the $k_{in}$ is concentrated on one value only;  the same
happens for $k_{out}$. For instance, this happens on lattices or
regular grids. Apart from being quite unlikely as cross-shareholding
configuration, empirical evidences in literature assess the power
law for the probability of $k_{out}$. Moreover, there is evidence
also on a power law or exponential behavior for the probability
distribution of $k_{in}$, as it is going to be detailed in the next
section. Such distributions are discrete and on a limited range of
integer numbers. In principle, these shapes of the marginal
distributions of the in- and out-degrees should prevent the
achievement of the minimum of the entropy, of course unless  the
joint structure is not the independent, but an ad-hoc one. It could
also happen that -- although keeping the power law/exponential form
-- the measures are so concentrated on their center of mass that the
entropy is quite close to its minimum. In this case, most of the
network units should have just one incoming and one outgoing link;
that is, again, a very unlikely configuration for a
cross-shareholding network. On the opposite, the maximum level of
 concentration increases when there is a flat uniform
distribution. In this case -- in order to make an example -- again
under the hypothesis of independence -- the units with the minimum
$k_{in}$ should have the maximal $k_{out}$; and vice versa (see the
Appendix 1 for further insights). This situation is much closer to
the kind of networks modeling the presence of mixed categories of
companies. In fact, usually financial companies land money in
exchange of shares; but sell their shares to a minimal number of
other companies, maximum one or two \cite{GRAMD}. On the opposite, manufactures sell their shares, but
rarely make financial investments buying shares of other companies -
unless strategically relevant to their specific
business \cite{GRAMD}.
\newline
In front of such different landscapes, some main research question
addressed in the present paper is exactly on these topics: is the
hypothesis of independence holding on a case study?  Is the network
topology of the case study limited to  the distribution of $k_{in}$
and $k_{out}$ sufficient, in itself, to prevent a rise of
concentration? Would there be maxima/minima of the entropy if -
keeping the marginals - the joint structure would be different?
 To which extent may the parameters describing the marginals
 change before eventually reaching maximum or minimum of concentration?

In order to achieve the tasks,
 we 
adopt a copula approach for assessing the concentration
 of the market through the stochastic dependence between in- and
 out-degree.
In this respect, copulas are of great usefulness (see
\cite{joe,nelsen}). Indeed, the classical Sklar's Theorem
\cite{sklar} explains that a copula function is able to represent
the connection between the joint probability distribution of a
random
 vector and the marginals of its components. Specifically, a
 multivariate copula computed over the marginals is equivalent to the
 joint distribution. Sklar's Theorem can also be read under a
 different perspective: starting from a joint distribution of a
 random vector and the marginals of its components, one can implement
 a best fit procedure to identify the copula describing the
 connection among them.

Thus, as already stated above, concentration is here captured
through the joint analysis of diversification and integration
 at an aggregate level.
Specifically, it is given by the Shannon entropy of the joint
distribution of in- and out-degree.  This leads to gain
insights on the market structure and on other relevant aspects, like
the reaction of the system to external shocks. Indeed, a polarized
market (minimum value of the entropy) can be associated to the
presence of a company with a central role, while a large entropy
suggests a fair distribution of the business network in terms of
companies ownerships.

It is worth remarking that a proper consideration of the
 weights of the network would make entropy equivalent to the Herfindahl-Hirschman (HH)  measure of
concentration, that became quite popular in financial studies after
its appearance in the official documents of the US mergers
guidelines for fixing alert threshold \cite{US}.

The present study offers to the regulatory bodies the possibility to
monitor the possible rise of concentration already looking to the
network topology only

For what concerns the dependence structure of diversification and
integration, we proceed under two different perspectives. By one
 side, we consider the independence copula and the Frechet bounds \cite{frechet},
 which are specific fundamental nonparametric copulas, and assume
 that they describe the dependence between the two degrees random
 variables. On the other hand, we calibrate the parameters of three
 families of copulas -- Gumbel, Clayton and Frank,
 see \cite{clayton,frank,gumbel}, respectively -- which belong to the classical
 family of Archimedean copulas \cite{ling}.

 In so doing, we focus on the informative content of the stochastic
dependence between in- and out-degree random variables. In fact, the
different copulas capture different
 stochastic dependence among the involved random variables.
In particular, Frechet bounds have an intuitive interpretation in
the bivariate case: they represent the maximum absolute values of
joint correlations. The upper bound stands for the highest positive
correlations, while the lower one is for negative correlations. The
Gumbel copula captures tail dependence, with a special attention
towards the dependence on the right tail. Differently, The Clayton
copula \cite{clayton} describes the dependence on the left tail of
the distribution. Frank copula \cite{frank} does not exhibit tail
dependence and allows both positive and negative dependence.

The methodology used for the calibration procedure is based on
 two different optimization problems, i.e. a maximum- and
 minimum-entropy for the joint distribution. In the former case, we
 are in the corner situation of an economic system with companies
 having the same values of diversification and integration; the
 latter case is associated to the maximum level of polarization, with
 only one company holding the total amount of connections, so that the maximum level of diversification and
 integration.

 In the same light, entropy is also computed in the case of
 nonparametric copulas for the obtained multivariate joint
 distribution. The paradigmatic cases of independence --
 product copula -- and maximum/minimum level of positive dependence
 -- the Frechet bounds -- serve as benchmarks.

The analysis has been also expanded for including a generic economic
system. Indeed, many empirical papers evidenced that the distribution of
the out-degree of many economic-financial systems is of a power law type \cite{Caldarelli}. Thus, the analysis has been replicated by
substituting the out-degree index with a power law function. The
parameter of the power law has been included in the set of
parameters to be calibrated. The empirical evidences on both the
existence of power law and of the exponential distribution for the
in-degree will be examined as well.

The generalization of the results of this paper to other
kind of networks, like networks with missing links is challenging
and useful. We have in mind contributions on not fully observable
networks that can be effectively adopted (see e.g.
\cite{Axel,Cimini,Serri}); this topic might be some matter for
future work.

%

The rest of the paper is organized as follows: the next sections
describe the selection of the probability distribution of the
marginals according to the existing literature and empirical data.
Section 3 presents the employed dataset. Section 4 outlines the
investigation procedure along with the considered copulas. Section 5
contains the obtained empirical results on the case study and on the
generalizations and discusses them. Last section concludes. Some
important ancillary results and materials are relegated in two
devoted Appendices.

\section{Distribution of the in- and out-degrees: empirical evidences in literature and a case study}
This section serves to fix the hypotheses on the shapes of the
marginal distribution that are meaningful for the problem under
examination.

In literature - most in the Econophysics realm - there was much
emphasis in the detection of the Pareto distribution in Economics
\cite{Feller}. Such a distribution is characterized by a power law
decay in the tails:
\begin{equation}
p(k) \sim k^{-\gamma}
\end{equation}
that corresponds to the cumulative distribution
\begin{equation}
P(k) \sim k^{1-\gamma}
\end{equation}

Therefore, if $k$ follows a power law with the exponent $-\gamma$,
then the cumulative distribution function $P(k)$ follows the power
law
with exponent $-\gamma +1$. 

\subsection{The out-degree $k_{out}$}
The presence of the power law in the distribution of the out-degree
is widely assessed in existing literature.

For example, Aoyama et al. \cite{Soumab,SoumaBook} add evidences to
the power law of the out-degree analyzing the shareholding network
of Japanese companies listed  in the Japanese stock market by using
only major shareholder data,  and focusing on companies concerned
with automobile manufacture. The results reported (see Fig. 4.28 and
Table 4.5 in  \cite{Soumab})  show the analysis of the
cumulative distribution of outgoing degrees in 1985, 1990, 1995,
2000, 2002, and 2003.  The size of the dataset ranges from $2078$ to
$3770$ companies, and all annual cumulative distributions can be
well fitted by a power-law distribution with exponents in the range
$(1.67,1.86)$, that leads to $\gamma \in (2.67,2.86)$

Souma et al. \cite{Souma} examine the Japanese shareholding network
existing at the end of March 2002. The network is constructed from
2303 listed companies and 53 non listed financial institutions. The
distribution of outgoing degrees is well explained by the power law
function with an exponential tail. 
The best fit of the cumulative is a power law with exponent $1.7$,
that corresponds to $\gamma= 2.7$.

In \cite{Garlaschelli} the direction of links reversal to the one
used in \cite{Bellenzier,Derrico,GRAMD}  is used for dealing with
diversification and integration, so their results for $k_{in}$
actually have to be compared with $k_{out}$ of the other papers. The
authors report also the power law exponents of some shareholding
networks: the Italian stock market (Milano Italia Borsa; MIB), the
New York Stock Exchange (NYSE), and the National Association of
Security Dealers Automated Quotations (NASDAQ). They find that all
of them follow a power law distribution: $\gamma_{MIB} =2.97$ in
2002, $\gamma_{NYSE} =2.37$ in 2000 $\gamma_{NASDAQ} =2.22$ in 2000.

The scale free structure has been estimated also  on the
shareholding of  223 companies quoted in MIB (Milan Stock Exchange)
in the time span 1/1/2004, 12/31/2004 \cite{Derrico}. Companies are
the network nodes; arcs are drawn from the shareholders to the owned
companies. The power law function with exponent $1.39$, that leads
to $\gamma=2.39$  nicely fits the distribution.

In \cite{GRAMD} the shareholding network of MIB companies are still
built as in \cite{Derrico}, but on data sampled in 2008. A best fit
estimate of $2.15$ and a Maximum Likelihood Estimate of
$\gamma=2.7$,  are in line with the above mentioned results.

In \cite{Chang}   the cross-shareholding of 300 index companies from
2007 to 2013 are studied. The companies are listed in   the Shanghai
and Shenzhen stock market. Data are  provided by the Securities
Times (STCN) and the Wind Database. The  sample of firms covers
about sixty percent of the market value of the Shanghai and Shenzhen
stock market. They find the following values of $\gamma$:
$\gamma=2.311$ (2007),  $\gamma=2.465$ (2008), $\gamma=2.558$
(2009),  $\gamma=2.625$ (2010),  $\gamma=2.721$ (2011),
$\gamma=2.722$ (2012),  $\gamma=2.724$ (2013).

In \cite{Li} the worldwide  network of listed energy companies
sampled in 2013 is built. The data source is the ORISE publicly
listed companies worldwide (https://osiris.bvdinfo.com), on December
31, 2013. There are 2334 listed energy companies and 8302
shareholders in the database (after removing duplicate items). In
this so large database, the power law exponent estimated for the
cumulative distribution of the out-degree is $\gamma=2.428$.


In \cite{Ma} the cross-shareholding networks of the companies listed
in Chinese stock market between 2002 and 2009 are studied. They
analyze the mutual investment at company-level, province-level and
region-level. However, they go beyond the mere topology of the
network, because they consider the weight of cross-ownerships into
the out-degree. Although they measure a quantity different from the
$k_{out}$ that we use in this paper, it is worth remarking that they
measure the power law in the range $(1.813-2.229)$\footnote{2.229
    (2002), 2.152 (2003), 2.057 (2004, 1.958 (2005),
    1.899 (2006), 1.788 (2007), 1.793 (2008), 1.813 (2009)}


%
%
%

The topological properties and evolution of the cross-shareholding
networks of listed companies Shanghai stock exchange and the
Shenzhen stock exchange  in China from 2007 to 2011 are analyzed in
\cite{LiAn}. They find that both the in-degree and the out-degree
follow a power law distribution in the range $(2.01, 2.43)$. In
detail: 2.43 (2007),  2.39  (2008), 2.33 (2009), 2.32 (2010),  2.33
(2011).

Vitali et al. \cite{Vitali} worked on the Orbis 2007 marketing database, that
comprises about 37 million economic actors, both physical
persons and firms located in 194 countries, and roughly 13 million directed and weighted
ownership links (equity relations). On such data, the power-law exponent of the
probability density function of the out-degree is   $\gamma=2.15$.

We may conclude that above empirical analyses allow to conclude that
the power law behavior of $k_{out}$ is quite widespread, and allows
us to assume a power law as hypothesis for $k_{out}$.

\subsection{The in-degree $k_{in}$}
The amount of empirical analyses of $k_{in}$ is much lower than the
ones on $k_{out}$. Some authors explicitly declare that they are not
interested in examining $k_{in}$, because the range of this variable
is more limited than $k_{out}$. A very few studies are available. In
\cite{Derrico} the in-degree distribution shows a power law, with
exponent $0.62$.  On \cite{GRAMD} data, the exponential distribution
was detected as the best fitting one, although the power law is
quite close. Therefore, we are going to examine both the power law
and the exponential as probabilities suitable for describing
$k_{in}$.

%



\section{Data}
The data is the set of holdings among listed firms in the Milan
Stock Market. It is the same as in \cite{GRAMD}. The data set has
been sampled on May 10th, 2008, from which we build the network of
shareholders and subsidiaries of companies traded on the MTA segment

{$www.borsaitaliana.it/azioni/mercati/mta/.../mta-mercato-telematico-azionario.en.htm$}

 of the Italian Stock Market. The
information available on several databases were cross-checked: the
Bureau Van Dijk databases and CONSOB for the active and passive
ownership sample; Bankscope for banking and financial companies;
ISIS for insurance companies; AIDA for all the remaining sectors;
Datastream Thomson Financial Database. The few companies that had
incomplete data on either active or passive holdings were excluded
from the present analysis. Even if very limited holdings (below 2\%)
have been considered, the mediate possessions held via mutual funds
were excluded as well, because they do not represent a direct
interest of a company into another.

The total size of the sample amounts to 247 companies, that
represent the nodes of the network, that is the 94\% of the total
number of listed companies and 95.22\% in terms of capitalization.
This dataset is slightly different from the one examined in
Garlaschelli et al. (2005) because some companies traded in the
market changed; moreover, there is a different level of accuracy in
the details of ownership data, and their $K_{in}$ corresponds to our $K_{out}$. Our notation for $k_{out}$ is following \cite{Chapelle}.

Most companies do not  actually buy shares of other companies, they
can be considered small companies. The giant component is made by
101 nodes, which are connected to each other \cite{GRAMD}. In the
present analysis, we consider only the values of the in-degree and
of the out-degree that are different from 0, so that we exclude
isolated nodes. The latter constitute the set of companies that do not buy shares of  (and which shares are not owned by) other companies traded in the same market.

\section{Investigation procedure}

This section is devoted to the introduction of the analytical
instruments used and to the description of the implemented analysis.

\subsection{The adopted copulas}
We firstly present the definition of bivariate copula, which is
crucial for the study.

A bivariate copula is a function $C:[0,1]^2 \rightarrow [0,1]$ such
that
\begin{itemize}
\item $C(u,v)=0$ if $u \times v=0$;
\item $C(u,1)=u$ and $C(1,v)=v$, for each $u,v\in [0,1]$;
\item Given the $2$-dimensional rectangle $[a_1,b_1] \times  [a_2,b_2]
\subseteq [0,1]^2$, then
$$\sum_{i_1=1}^2 \sum_{i_2=1}^2 (-1)^{i_1 +i_2} C(u_{i_1},v_{i_2})\geq 0,$$
where $u_{j}=a_j$ and $v_{j}=b_j$.
\end{itemize}

The concept of bivariate copula plays a key role in describing the
stochastic dependence between two random quantities. Such a
statement is formalized in the Sklar (1959)'s Theorem, reported
below:

{\bf{Theorem}}\label{theorem:sklar}\\
Let $P$ be the joint distribution function of a bivariate random
variable $(X,Y)$. Define the margins as $P_{X}$ and $P_{Y}$. Then
there exists a bivariate copula $C$ such that, for each $(x,y) \in  {  R}^2$,
\begin{equation}
\label{eq:Sklar} P(x,y)=C(P_{X}(x), P_{Y}(y)).
\end{equation}
If the margins $P_{X}, P_{Y}$ are continuous, then the copula $C$ is
unique. Conversely, if $C$ is a bivariate copula and $P_{X}, P_{Y}$
are distribution functions, then the function $P$ defined in
(\ref{eq:Sklar}) is a bidimensional distribution function with
margins $P_{X},P_{Y}$.

Theorem \ref{theorem:sklar} explains that the relationship between
the joint and the marginal distributions of a couple of random
variables can be formalized by employing copulas.

Different copulas describe different types of stochastic dependence.
The analysis here implemented refers to six copulas --- or classes
of copulas --- which are widely used in the applications.

Specifically:
\begin{itemize}
\item Product copula
\begin{equation}
\label{ind} C_I(u,v)=uv.
\end{equation}
This is the case in which the random variables $X$ and $Y$ are
independent.

\item Lower Frechet bound
\begin{equation}
\label{LF} C_{LF}(u,v)=\max\{u+v-1,0\}
\end{equation}
This copula represents the case of perfect negative correlation
between $X$ and $Y$.

\item Upper Frechet bound
\begin{equation}
\label{UF} C_{UF}(u,v)=\min\{u,v\}
\end{equation}
This copula, in an opposite way with respect to the previous one,
captures perfect positive correlation between $X$ and $Y$.

\item Gumbel Archimedean copula
\begin{equation}
\label{G} C_G(u,v)=\exp[-((-\ln (u))^\theta+ (-\ln
(v))^\theta)^{1/\theta}], \qquad \theta \in [1, +\infty)
\end{equation}
In this case, one has an asymmetric tail dependence, with more mass on the right tail. Such a dependence is influenced by the value of
the parameter $\theta$.

\item Clayton Archimedean copula
\begin{equation}
\label{C} C_{C}(u,v)=\left[\max\{u^{-\theta}+v^{-\theta}-1,0\}
\right]^{-1/\theta}, \qquad \theta \in [-1, 0)\cup (0,+\infty)
\end{equation}
Analogously to the previous case, here one has an asymmetric tail
dependence . However, Clayton copula is associated to a predominance
of the left tail.

\item Frank Archimedean copula
\begin{equation}
\label{F} C_{F}(u,v)=-\frac{1}{\theta}\ln\left[1+\frac{(\exp(-\theta
u)-1)(\exp(-\theta v)-1)}{\exp(-\theta)-1} \right], \qquad \theta
\not=0
\end{equation}
This copula is not associated to tail dependence, and is  able to
capture either positive or negative dependence on the basis of the
value of $\theta$.

\end{itemize}

Product copula and the Frechet bounds are associated to
nonparametric functions, since they do not depend on any parameter.
Differently, the presence of a scalar $\theta$ in the definition of
Gumbel, Clayton and Frank copula says that such copulas are of
parametric type.

\subsection{Outline of the analysis and numerical results} The
availability of the case study allows to have a full description of
the marginals and of the joint distribution of the in- and
out-degrees. However, the general case is also included for the sake
of universality of the analysis.

The investigation procedure is split in three cases. In all
the steps, the above-mentioned copulas are taken as reference
instruments, in order to describe stochastic dependence between the
in- and the out-degree and achieve different objectives.

In the case 1, a description the empirical data coming out
from the available sample is provided. Starting from the empirical
(marginal) distributions of in-degree and out-degree, we derive the
joint distribution of such quantities by applying Sklar (1959)'s
Theorem through the copulas
introduced above. The Euclidean distance between the non-parametric
copula-based distributions are computed, and also the calibration of
the parameters of the Archimedean copulas are obtained by a
Euclidean distance minimization.

Case 2 still focuses on the case study. Substantially, this
step can be viewed as a replication of the previous one with the
remarkable difference that the Euclidean distance has been replaced
by the Shannon entropy. The meaning of this second step of the
analysis can be easily synthesized. Indeed, we here look at the
conditions on the stochastic dependence between in- and out-degrees
leading to market polarization (minimal entropy) or market fairness
(maximal entropy). In so doing, we derive information on the way in
which the degrees should be connected in order to shape the market.
Two separate cases are treated: first, computation of the entropy
for the cases of non-parametric copulas; second, the calibration of
the parameters of the considered Archimedean copulas under a
maximum- and minimum-entropy approach.

In the case 3, we provide a generalization and, in accord to
the existing literature, we consider marginal densities depending on
parameters. In details, we consider power-law and exponential for
the out-degree, while we take the in-degree without parametrization,
according to its empirical distribution. Also in this case, two
cases are treated: first, the non-parametric copulas are imposed and
the parameters of the power laws and exponential are calibrated
under a maximum- and minimum-entropy approach; second, the
parametric copulas of Gumbel, Frank and Clayton types are considered
and their parameters, along with that of the out-degree
distribution, are calibrated in a max/min entropy approach.

The probability of configuration $P(k_{in}=i,k_{out}=j)$ is
calculated through the copula as
$P(k_{in}=i,k_{out}=j)=C(u(i),v(j))-C(u(i-1),v(j))-C(u(i),v(j-1))+C(u(i-1),v(j-1))$.

Moreover, the calibration methods might naturally be based
on other concepts of distance (see e.g. \cite{liese,rachev}). In
this respect, it is also worth mentioning the results and
methodologies proposed in Schellcase (2012), where the author provides an
estimation of copula density through penalized splines of different
types \cite{schellcase}. However, as already pointed out above, Euclidean distance and
entropy have different meanings and are particularly suitable for
capturing the focuses of our investigation purposes.

\section{Results and discussion}
The obtained findings of the analysis are here described and
discussed.

\subsection{Case 1: distance from the empirical joint distribution}

\begin{figure}
    \centering
    \includegraphics[width=1\textwidth]{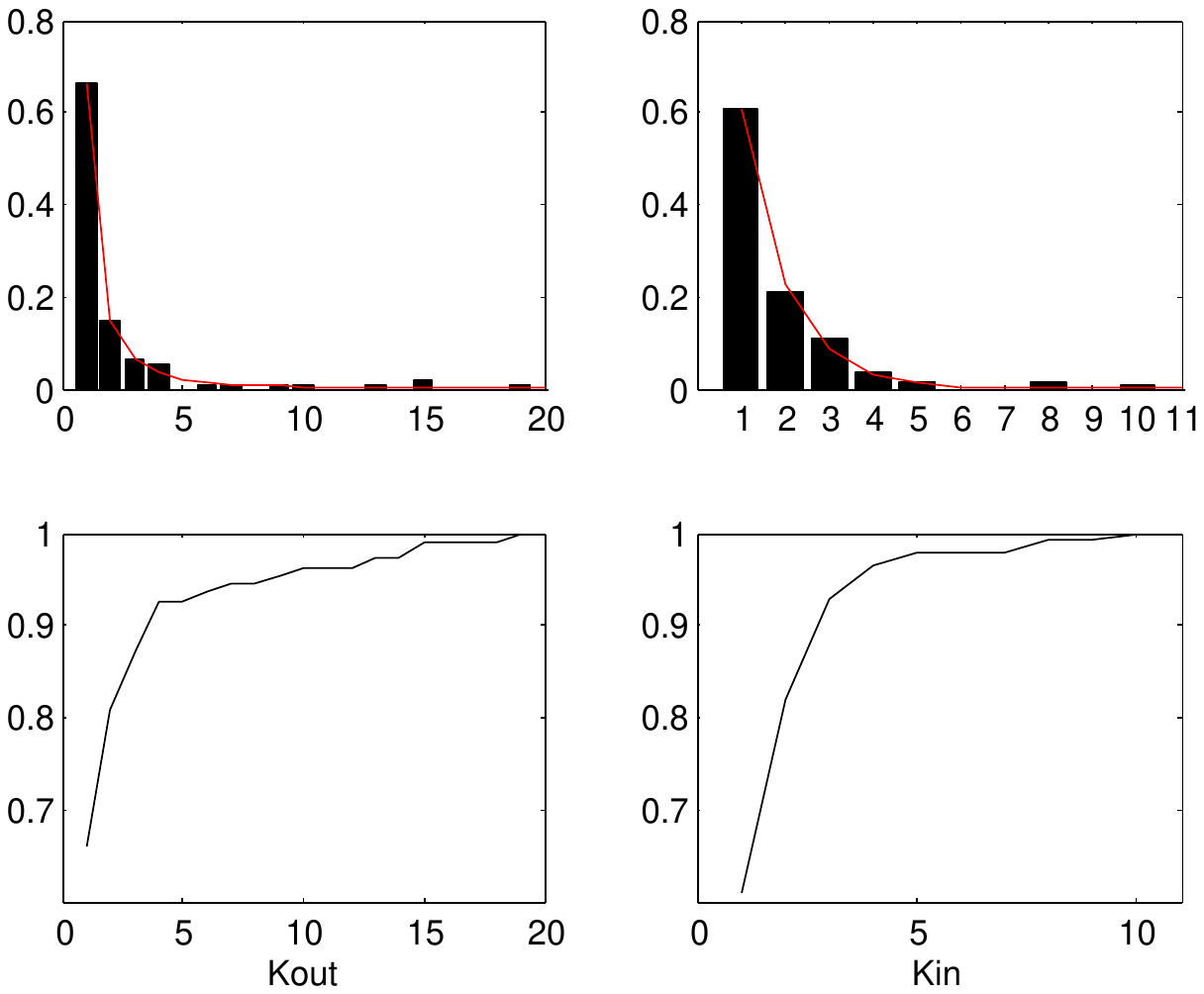}
    \caption{
Upper figures: histograms (empirical densities, left:
$p(k_{out}=x)$, right: $p(k_{in}=x)$). Lower figures: distributions
(left: $P(k_{out}<x)$, right: $P(k_{in}<x)$). The left part
corresponds to Fig. 4 of \cite{GRAMD}.} \label{fig:CaseStudyMarginal}
\end{figure}

\begin{figure}
    \centering
    \includegraphics[width=1\textwidth]{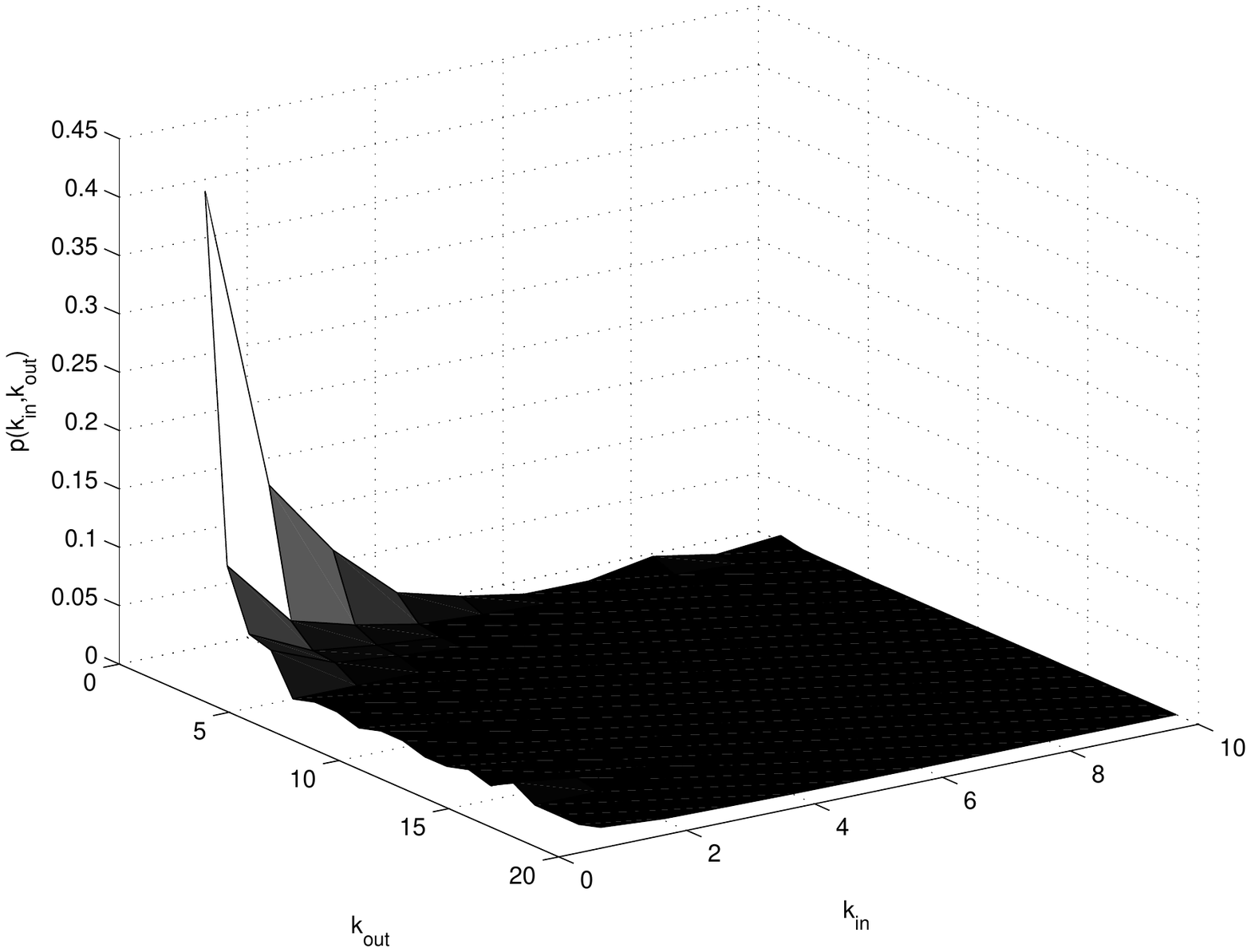}
    \caption{Case study. Joint empirical distribution.}
\label{fig:CaseStudyJoint}
\end{figure}

Figure \ref{fig:CaseStudyMarginal} shows the empirical marginal
distribution of $k_{in}$ and $k_{out}$ for the empirical case we
deal with, while Figure \ref{fig:CaseStudyJoint} shows the joint
probability. The range for $k_{in}$ is $[1, \cdots, 10]$, and
 the range for $k_{out}$ is $[1, \cdots, 19]$.
The limits of $10$ for $i$ and $19$ for $j$ are due to the specific
sample. The value $0$ is not considered in the present analysis. In
fact, the detection of the Pareto  distribution would mainly concern
the tails. Thus, we notice that there are too many $0$'s for
appreciating such a distribution in the full histogram.

The power law best fit over the density gives $p(k_{out}) \sim
k_{out}^{-\gamma}$ with $\gamma = 2.159(1.984, 2.339)$, RMSE=0.0094.

The Jarque-Bera test validates the hypothesis of Gaussianity of
residuals. The power law best fit on the empirical probability
distribution leads to $ P(k_{out}) \sim k^{1-\gamma} $ where  $\gamma =
1.7925 (1.6596, 1.9254)$, RMSE=0.0088. The MLE $\gamma$ gives
$\gamma = 2.72766 (2.72763, 2.72768)$.
%
%
For the case of in-degree, the Jarque-Bera test rejects the
hypothesis of Gaussianity of residuals. Therefore, there is still
residual information in the residuals whence the hypothesis of power
law decay cannot be fully validated. However, the empirical
distribution is quite close to the power law. For the in-degree
$k_{in}$ the best fit is the exponential General model Exp1: $f(x) =
a*exp(b*x)$ Coefficients (with 95\% confidence bounds): a = 1.6
(1.424, 1.777) b = -0.9727  (-1.061, -0.8845) Goodness of fit: SSE:
0.001137 R-square: 0.9966 Adjusted R-square: 0.9963 RMSE: 0.01124.


The parametric copula --- Gumbel, Frank and Clayton --- that best
fits to the empirical data is now detected. For the non parametric
copulas we calculate the distance $d(C_I,P)$ of the joint
distribution calculated by using the copula $C(u,v)$ from the
empirical joint distribution $P$. Such a distance will be used as a
benchmark value.

The results are:

\begin{itemize}
\item Product copula (independence):  $d(C_I,P)=4.06e-014$

\item Lower Frechet bound $d(C_{LF},P)=0.9354$

\item Upper Frechet bound $d(C_{UF},P)=3.9484$
\end{itemize}

Therefore, the joint empirical distribution is closer to the
hypothesis of independence (product copula) than to the others.

\begin{figure}
    \centering
    \includegraphics[width=1\textwidth]{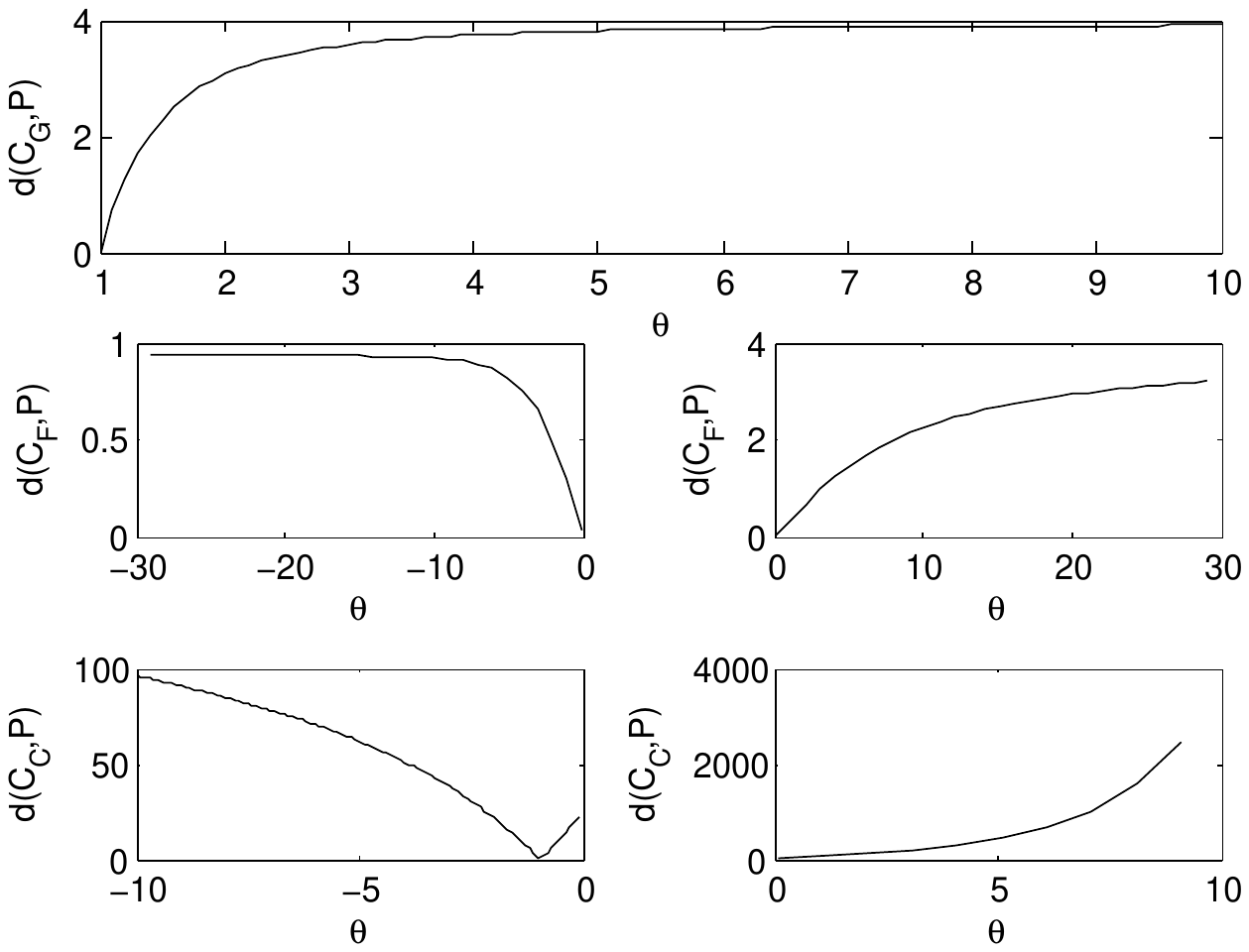}
\caption{Distance $d(C,P)$ from the empirical distribution, when the
joint distribution is calculated through the Gumbel (upper figure,
$d(C_G,P)$), Frank (middle figures, $d(C_F,P)$) or Clayton
distribution (lower figures, $d(C_C,P)$).}
\label{fig:distancesEmpirical}
\end{figure}

On the copulas that depend on a parameter a best fit procedure has
been implemented. Figure \ref{fig:distancesEmpirical} plots the
dependence of the distance on $\theta$  considering the three cases
for the joint distribution: the Gumbel, Frank and Clayton copulas:

\begin{itemize}

    \item Gumbel Archimedean copula. The best fit holds for  $\theta=1$, with
practically $0$ as value for the distance. This is coherent with the
case of the product copula, because, in fact, when $\theta=1$, then
the Gumbel copula reduces to the product copula. Small differences
on the distance are due to the numerical rounding of the algorithm.
This outcome confirms what obtained for the independence case.

\item Frank Archimedean copula. The distance from the empirical data is decreasing as
$\theta$ approaches 0, but $0$ does not belong to the definition
set. Therefore, the calibrated parameter tends to zero. We do not
have an optimal value of $\theta$. From this, we infer that this
copula is not suitable for the fit.

\item Clayton Archimedean copula.
For the negative values of $\theta$, there is a minimum for
$\theta=-1$, that belongs to the definition set and corresponds to
the case of the lower Frechet bound. The value of the distance for
$\theta=-1$ is $ 0.93$.

\end{itemize}

Thus, the empirical in- and out-degrees exhibit a structure
of stochastic independence, with a very small value of the distance
between the empirical distribution and the one obtained in the
product copula case. This is also confirmed in the Gumbel copula
case. However, when forced to describe a type of dependence
described through a Clayton copula, data are less distant from an
absolute negative correlation (lower Frechet bound). This outcome is
in agreement with the fact that the distance of the data from the
lower Frechet bound is lower than the one from the upper Frechet
bound.
\newline
Under an economic point of view, independence means that there is
not a regular behavior of companies in the respect of integration
and diversification. More precisely, it is not possible to infer
diversification properties of the market by looking at the
integration, and vice versa. 

\subsection{Case 2: entropy}
In this section, we start working on the entropy. We refer to the
Shannon entropy \cite{Shannon}
\begin{equation} \label{entropia}
 H(C(u,v,\theta))=-\sum_{u,v} C(u,v,\theta)\ln C(u,v,\theta)
 \end{equation}
The entropy calculated on the empirical joint distribution is
$1.52$. On the joint distribution calculated through the copulas not
depending on parameters, the values of the entropy are:
\begin{itemize}
\item Product: $H=1.52$, the same value as for the
empirical joint distribution. In fact, this copula well describes
the joint distribution.
\item Lower Frechet: $H=0.96$.
\item Upper Frechet: $H= 1.45$.
\end{itemize}

\begin{figure}
    \centering
    \includegraphics[width=1\textwidth]{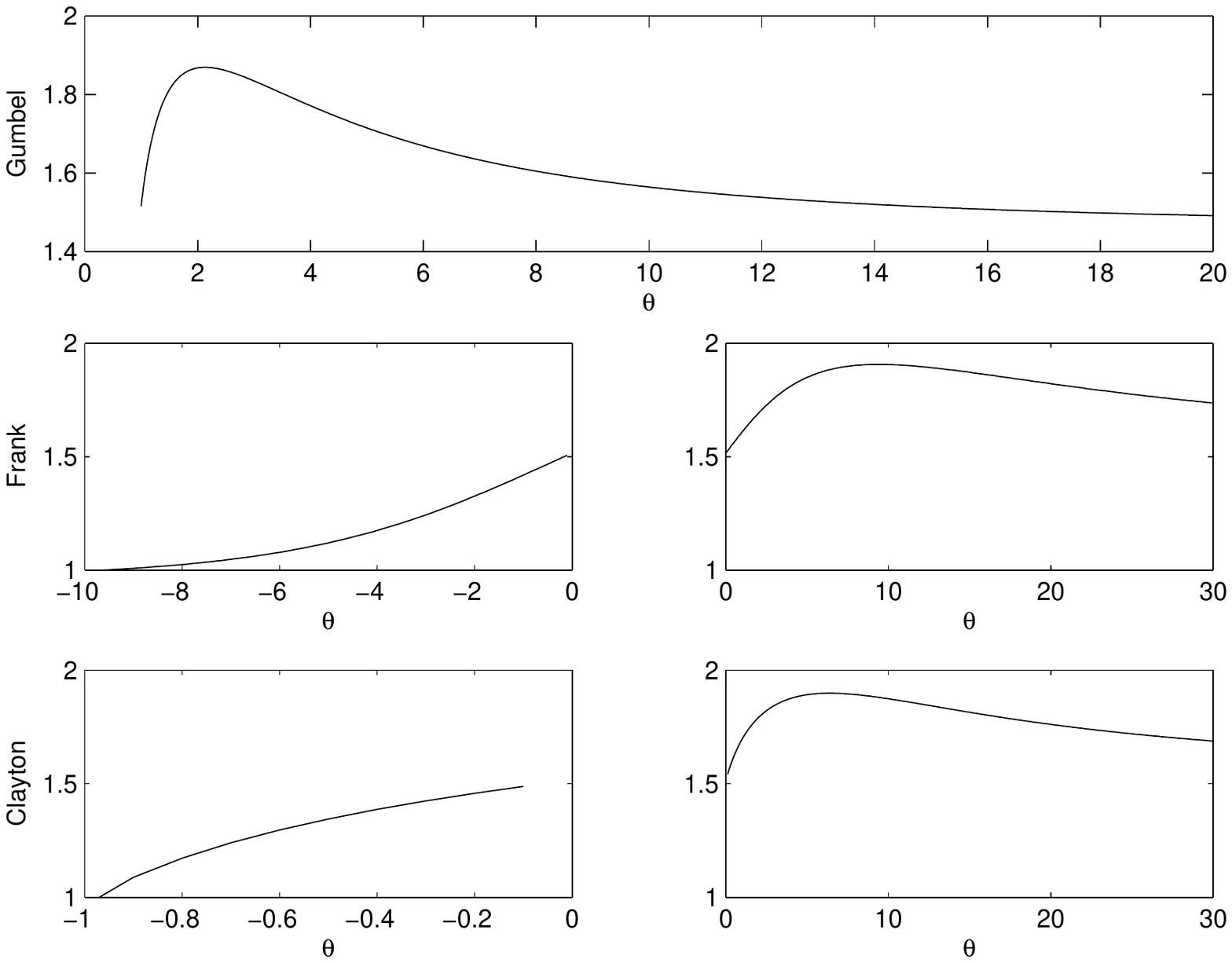}
    \caption{Plot of the dependence of the entropy function
on the parameter theta for the Gumbel, Frank and Clayton, calculated
on the marginals of the case study. Clearly, no minimum internal to
the definition sets. There is a maximum for the Gumbel copula  in
$\theta=2.13$. There is a maximum for the Frank copula  in
$\theta=9.41$. There is a maximum for the Clayton copula  in
$\theta=6.39$. }
        \label{fig:Hteta}
    \end{figure}

For the parametric copulas, we perform a comprehensive analysis on
the minimum/maximum as a function of  $\theta$. Figure
\ref{fig:Hteta} shows the dependence of the entropy on $\theta$ in
the cases of joint distribution calculated through copulas. We get
the following results:

\begin{itemize}

    \item Gumbel Archimedean copula.
The numerical minimization
procedure gives the best fit for $\theta= 1$, with a value of the
entropy equal to $1.5154$ . This is in line with the best fit of the
product copula. From Figure \ref{fig:Hteta} it is possible to note
that there is an asymptotic behavior for $\theta$ going to infinity.
The maximum is attained for $\theta=2.1312$ with a value of the entropy equal to $1.8693$.

    \item Frank Archimedean copula.
There is no minimum because $0$ does not belong to the definition
set of the functions. The maximum is attained for  $\theta=9.4205$
with a value of the entropy equal to $1.9060$.

\item Clayton Archimedean copula. There is no minimum internal
to the definition set. From Fig. \ref{fig:Hteta} it is clearly
visible that the function is decreasing for $\theta<0$, so
$\theta=-1$, that is the lower bound of the parameter variation
interval, is a point of minimum. Regarding the maximum, the
numerical maximization of the entropy gives the point of maximum in
$\theta=6.3899$, with a value of the entropy equal to $1.8982$.
\end{itemize}

Results can be commented as follows. Independence is
confirmed to describe the stochastic dependence between the degrees.
More than this, we can also say that data are associated to a high
value of the entropy. This outcome says that the market described by
the considered companies has a "broadly fair" distribution in terms of
integration and diversification. Such a "fairness" is more evident in
the cases of Frank and Clayton copulas, whose calibrated parameters
suggest that left tail dependence (Clayton) and positive correlation
(Frank) are more likely associated to a uniform distribution of the
in- and out-degrees. We point out that the left tail dependence is
related to the presence of a strong correlation when the levels of
diversification and integration are low.

The detection of a maximum shows that there are possible
configurations for the joint distribution that lead to a network
where the in-degree  (distribution) is decoupled from the out-degree
(distribution). Situation like this may happen when companies are
artificially created, so that a wide set of combinations is
possible: nodes with low (high) in-degree and high (low) out-degree
or nodes with  similar values of in-degree and out-degree. For
instance, in the MIB30 (\cite{GRAMD}, Figure 1) the company IFI PRIV
was created for controlling IFIL, that has the main role to provide
financial services to the main companies of the Agnelli family: FIAT
and JUVENTUS, so IFIL has only one outgoing link, and no incoming
links - the ultimate owners being the persons member of the family.
In \cite{GRAMD}, while Figure 2 in the quoted paper shows a list of
companies for which the only link is due to the need of using a
financial institution - that, in turn, gets ownership of the
financed company. A circumstance that leads to quite different
values for $k_{in}$ and $k_{out}$ for a single node is given by the
role of banks and  insurance companies: since they provide money to
other companies, they get in exchange the ownership, whence  having
many outgoing links. On the other side, they use insurance companies
transferring them their   own part of their risk. In \cite{GRAMD},
Figure 1, on the left, the cases of MPS bank and UNIPOL insurance
company clearly evidence this kind of situation.

%
%


\subsection{Case 3: marginals depending on parameters}
The previous section has shown the case study. In literature, most
often the $k_{out}$ follows a power law, with exponents in a range
$(2 , 3)$. The few studies on $k_{in}$ have shown most either a
power law or an exponential. In this section, we aim at extending
the previous results to a more general case in which the exponent of
the power law may change. This corresponds to study the effect of a
change of exponents on the results of the maximization and
minimization of the entropy. It is worth recalling that the exponent
of the power law has an implication on the presence of fair values.
The higher the exponent, the faster is the decrease, meaning that
there are many low values of the degrees and a very few with high
ones. For instance, in \cite{Derrico} the MIB30 network of
cross-shareholding was showing a power law. In fact, the companies
considered in the quoted paper were more keen to diversify their
investment.
The crisis in 2008 canceled this kind of
investment, as shown by the increase of the value of the power
law exponent on the MIB30 in 2008 \cite{GRAMD}.



Although the power law remains the best fitting, the shape of the
distribution is slowly moving to a sharply decreasing function,
becoming closer to an exponential distribution. The same behavior of
a distribution has been shown in \cite{Clementi} in the context of
wealth.

For each of the above listed copulas, we here look for the minimal
and maximal entropy using the following marginal distributions:
\begin{enumerate}
    \item step 1: power law for $k_{out}$, and raw data for $k_{in}$.
    \item step 2: raw data for  $k_{out}$, and power law for $k_{in}$.
    \item step 3: raw data for  $k_{out}$, and exponential law for $k_{in}$.
    \item step 4: power law for $k_{out}$, and power law for $k_{in}$.
    \item step 5: power law for $k_{out}$, and exponential law for $k_{in}$.
\end{enumerate}
The last two cases correspond to the most general case, independent
from the case study. For each of them, all the copulas listed in the
methodological section  are tested.

To be concise and informative, we present here only step 1.
The interested reader can find the other cases in Appendix B.

\subsubsection{Step 1:  power law for $k_{out}$, and raw data for $k_{in}$}

In this case we consider the cumulative distribution $P(k_{out}<x)=a
x^{-k+1}$. We are not considering the more general functional form
$a x^{-k+1}+b$ because the density in this kind of problems is
vanishing as $k$ increases, so $b$ would be $0$. The parameter $a$
is automatically fixed by the normalization condition
$P(k_{out}<\infty)=1$.

We already pointed out that the parameters regulate the mass
distribution over the range. Low values of $k$ lead to a more flat
distribution; high values of $k$ increase the skewness to the left,
and so the cumulative distribution function is quickly growing at
the beginning of the range; the inflectional point is moving to the
left. The increase of the skewness leads to an alignment to the
distribution of $k_{in}$, so increasing the peakness and the
concentration of the distribution, hence the minimization of the
entropy. Here below, we report results for both parametric and non parametric copulas.  The Figures referring to non-parametric copulas  report $k$ on
the $x$-axis for the non parametric copulas. The parametric copulas
depend on $k$ and $\theta$, but the 3D visualization is less clear than the 2D one. Therefore, the visualization  for the parametric copulas is more clear drawing the entropy as function of
$\theta$ (on the $x$-axis) for different meaningful values of $k$ (corresponding to different curves).
\begin{itemize}
\item Non parametric copulas: Fig. \ref{fig:NonParametricKinFisso}
shows the behavior of the entropy as a function of $k$. The upper
Frechet bound and the product copula are quite overlapped: the
entropy increases as $k$ increases. Practically, in the marginal of
$k_{out}$ the entropy is minimal as the  mass is pushed to the
highest mass concentration of $k_{in}$, that is at the left bound of
the domain, although it should not become more sharp than the
empirical distribution of $k_{in}$. This is coherent with the
Theorem in the Appendix, as well as with the very well known fact
that the entropy is minimal as the dispersion diminishes and the
mass is concentrated. The lower Frechet copula has the opposite
behavior. There is no minimum and no maximum internal to the range
for $k$. All the three show a maximum: for $k=0.46$ and $H=2.12$
(Product), $k=2.2$ and $H=0.97$ (Lower Frechet), $k=1$ and $H=2.09$
(Upper Frechet). The only maximum in the most interesting range of
$k \in (2,3)$ is the Upper Frechet one. In the Frechet one there is
also another local maximum in $k=0.81$ and $H= 0.52$ and two local
minima in $k=0.71$ and $H= 0.50$ and in $k=0.91$ and $H= 0.48$. The
other local fluctuations in the Upper Frechet do not lead to other
local maxima or minima. All the entropies are decreasing for $k$
increasing.
            \begin{figure}
                \centering
                \includegraphics[width=1\textwidth]{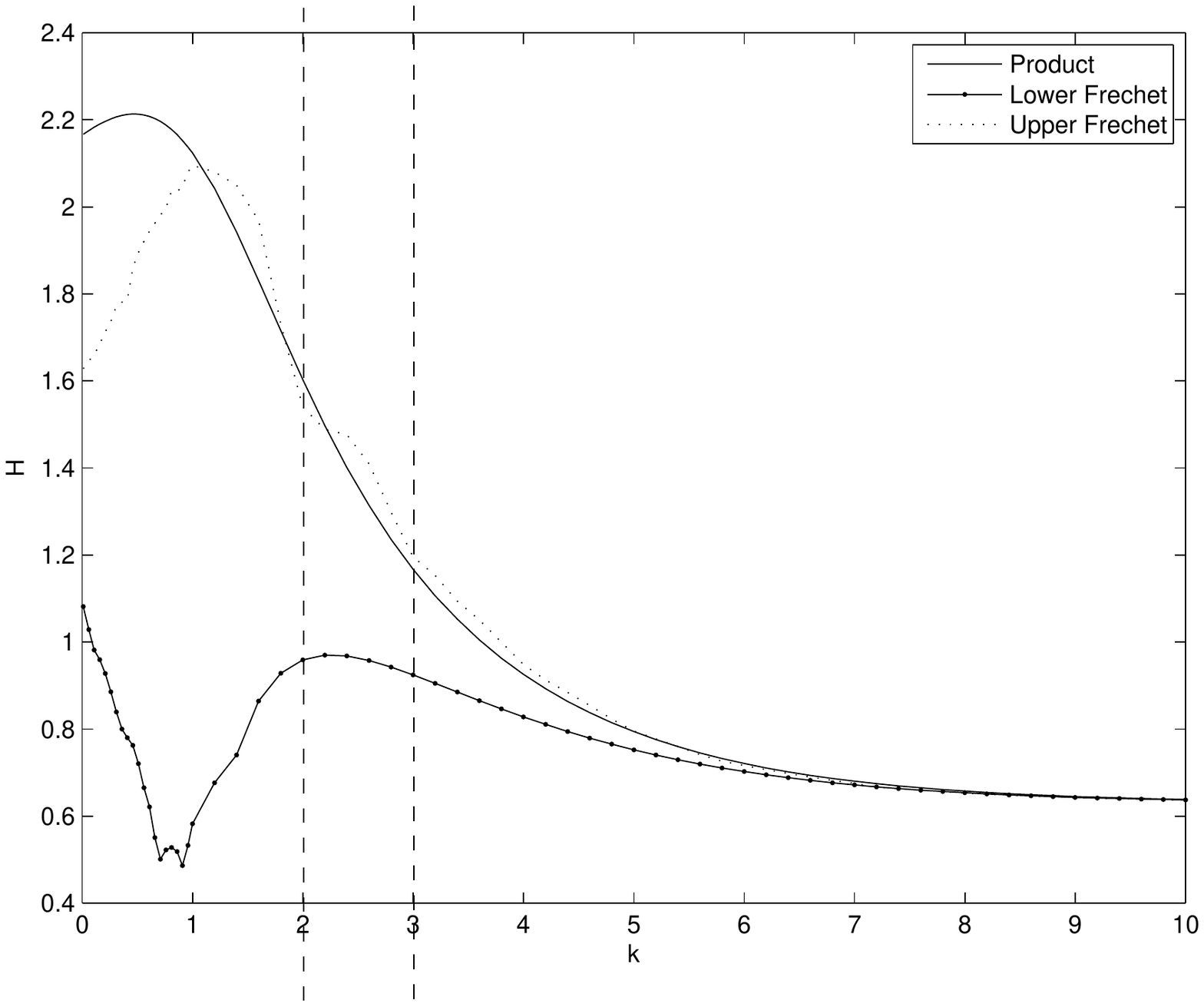}
\caption{The figure shows the dependence of the entropy on $k$ for
each of the three non parametric copulas.}
\label{fig:NonParametricKinFisso}
            \end{figure}

\item  Fig. \ref{fig:Prova2GumbelM0} shows the entropy
function when the exponent of the power law for $k_{out}$ is allowed to change.  Therefore, the marginal distribution is allowd to change, still remaining a power law. The other marginal  is
given by the case study for $k_{in}$. The marginal distributions are combined through  the
Gumbel copula. The minimum that was detected on the raw data  for $
\theta=1$ disappears, and an asymptotic behavior remains: the
entropy is decreasing for $ \theta \rightarrow \infty$, i.e. in the
case of convergence towards the Frechet upper bound. Therefore, the
minimum entropy is obtained either when the copula is the product or
when the considered quantities are perfectly positively correlated.

Once more, we may remark that the entropy decreases as the
concentration of the distribution increases, possibly reaching a
Dirac's delta function. Since the marginal on $k_{in}$ is fixed, the
minimum is obtained when the mass through the other marginal is
concentrated on the highest peak of $k_{in}$, that is at the left
border. This effect is obtained by increasing the steepness of the
marginal of $k_{out}$. The higher $k$, the more the mass is
concentrated on the left border. This effect is emphasized by the
application of the copula. Since both marginals are left-skewed, the
product gives the minimum, for quite a range of values of $k$.
However, the entropy is decreasing as $\theta \rightarrow \infty$,
reaching values lower than the minimum, when present. Therefore any
concentration limit can be overrun, providing that the slope of the
power law is large. We already noted that most systems show a power
law with an exponent between 2 and 3. This prevents the rise of
concentration. 
\newline The analysis of the maximum is quite different. As $k$
increases, the maximum is pushed to the left side of the range of
$\theta$, tending to  $1$ for high values of $k$, i.e. in the case
of independence.

        \begin{figure}
            \centering
            \includegraphics[width=1\textwidth]{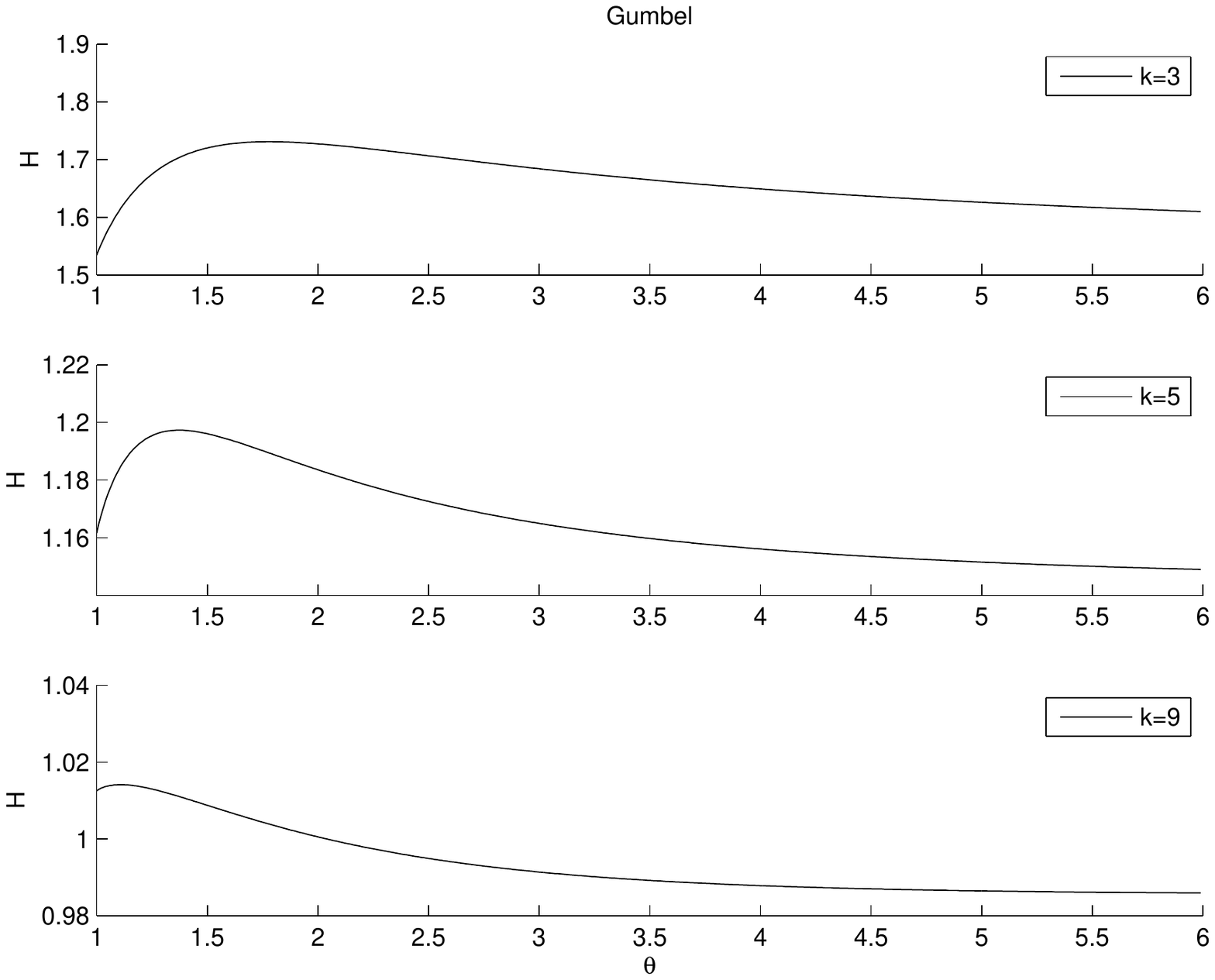}
\caption{The figure shows three cases for the entropy (y-axis) as a
function of $\theta$ (x-axis). The marginals are: the power law for
$k_{out}$ and from the case study for $k_{in}$. They are combined
through a Gumbel copula. In all   cases, the function is decreasing
as $\theta \rightarrow \infty$. The maximum is well evidenced, like
in our case study. As $k$ increases, the maximum moves to the left
border.  
} \label{fig:Prova2GumbelM0}
        \end{figure}

\item Frank copula. Also for the Frank copula there are different configurations as the parameters of the power law changes.
Figure \ref{fig:Prova2Frank} outlines the situation for $\theta <0$
(left hand side) and for $\theta >0$ (right hand side).

The Frank copula when $\theta<0$ gives a result similar to the left
part of the second row of the figure  \ref{fig:Hteta}: there is no
minimum. Moreover, the value of the entropy is increasing as
$\theta$ increases. However, for each fixed $\theta$, the values of
the entropy decreases as $k$ increases. If $\theta >0$ the maximum
moves to the right as $k$ increases. There is no minimum, since $0$
does not belong to the definition set.

        \begin{figure}
            \centering
            \includegraphics[width=1\textwidth]{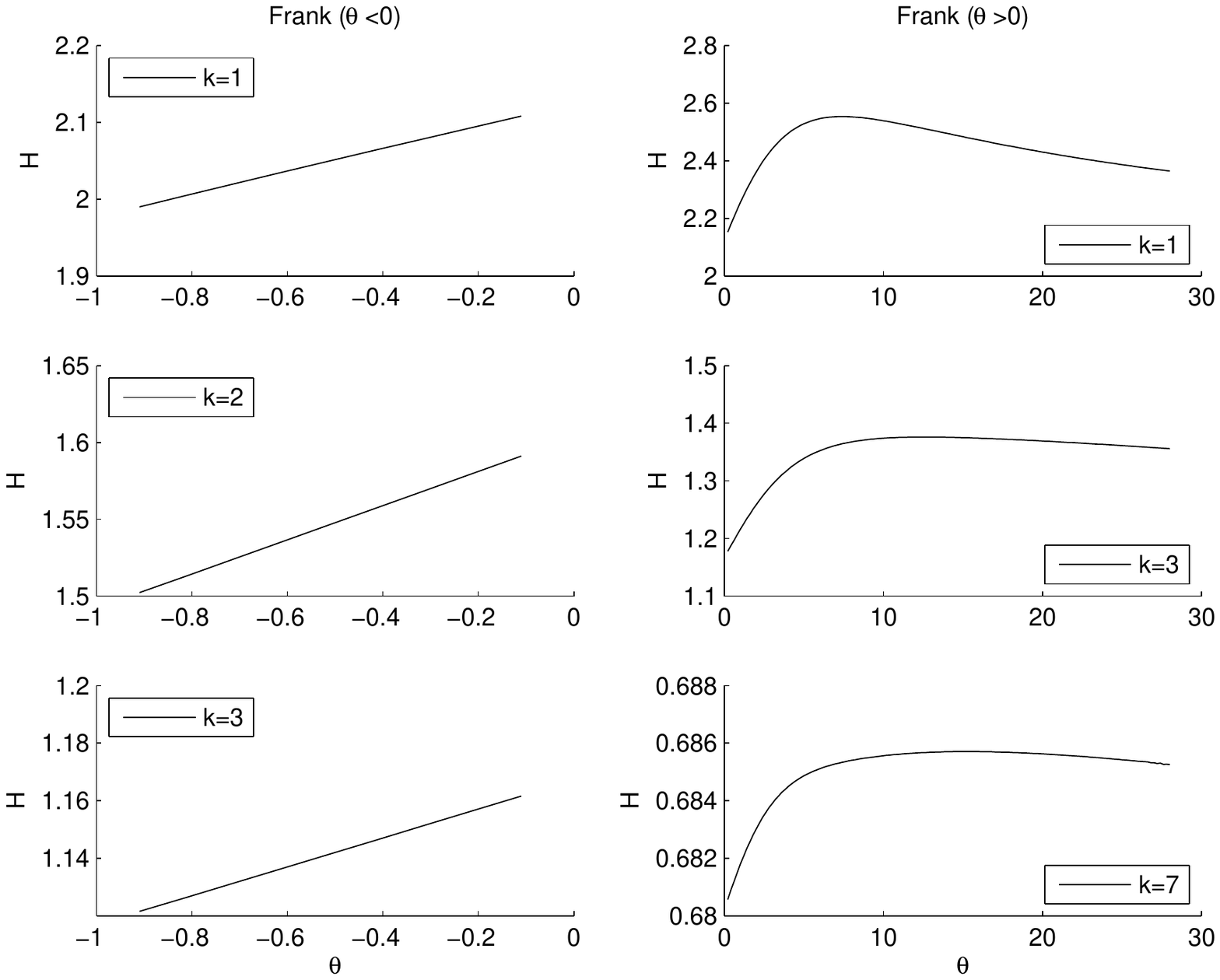}
\caption{The figure shows three cases for the entropy (y-axis) as a
function of $\theta$ (x-axis). The marginals are: the power law for
$k_{out}$ and  the empirical distribution of case study for
$k_{in}$. They are combined through a Frank copula. When  $\theta
>0$, the maximum moves to higher values of $\theta$ as $k$
increases. Since $0$ does not belong to the definition set, there is
no minimum. Left side of the figure: in all   cases the function
is increasing for $\theta \rightarrow 0^+$ and decreasing for
$\theta \rightarrow -\infty$.
            } \label{fig:Prova2Frank}
        \end{figure}

\item Clayton copula.
Figure \ref{fig:Prova2Clayton} shows the situation depending on the
parameters of the power law. For $\theta >0$, the subplots show that
the maximum moves to the right hand side as $k$ increases. There is
no minimum, since $0$ does not belong to the definition set, there
is no minimum. For $\theta <0$, there is a minimum for $\theta=-1$,
for any value of $k$.

        \begin{figure}
            \centering
            \includegraphics[width=1\textwidth]{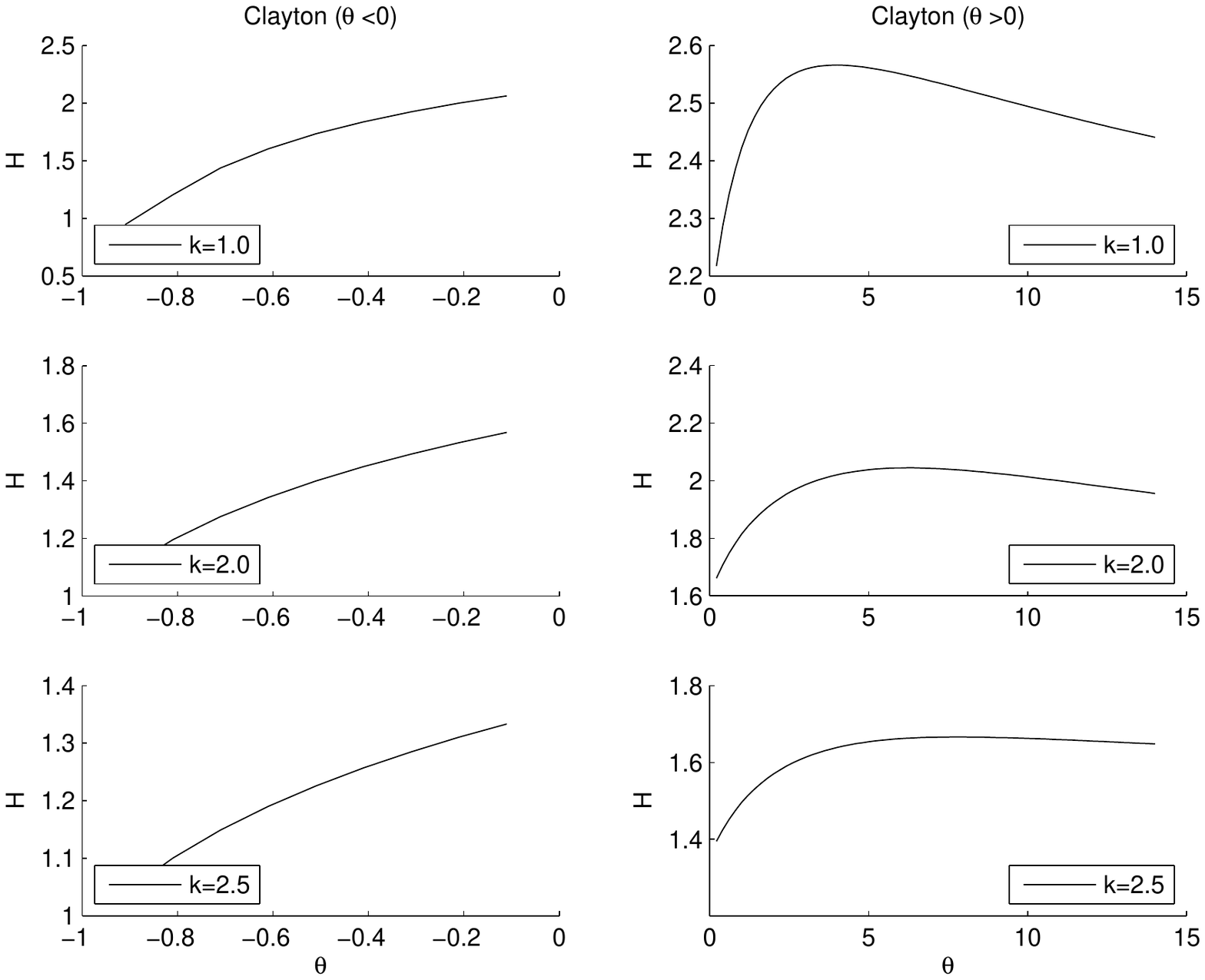}
\caption{The figure shows three cases for the entropy (y-axis) as a
function of $\theta$ (x-axis). The marginals are: the power law for
$k_{out}$ and from the case study for $k_{in}$. They are combined
through a Clayton copula with parameter $\theta <0$. The left
figures shows the case  $\theta <0$. There is a minimum for
$\theta=-1$, for any value of $k$. The right figures show the case
$\theta >0$. The maximum moves to the right hand side as $k$
increases. For any $k$, the function is decreasing for $\theta
\rightarrow -\infty$. } \label{fig:Prova2Clayton}
        \end{figure}

\end{itemize}

\section{Conclusions}

This paper provides a detailed analysis of the concentration of a
market, which is captured by a joint analysis of diversification
and integration. Such concepts are strongly linked with the network
described by the cross-shareholding matrix and the related
 entropy measure. In particular, the out-degree value of a company
formalizes its diversification while  the in-degree  value is
related to its integration in a network of shareholders. The
analysis of such degrees may be relevant for regulatory bodies, that
need to fix thresholds and eventually capture early signals for
preventing concentration. Literature studies have shown that the
most frequently detected probabilities for description of
diversification and integration were the power law and the
exponential law. The parameters of the distribution regulate their
shape. However, it is the coupling between in- and out-degrees which
is the most relevant to the concentration evolution.

The dependence between the components of the matrix --- the in- and
out-degrees --- is here captured through appropriately selected
copulas. Among them, the most prominent examples of nonparametric
copulas --- product and Frechet bounds --- are also included. The
maximum of concentration can be achieved by minimizing the entropy.
When one marginal distribution is fixed, the results show that  the
minimal entropy is achieved when the other  marginal distributions
gather at the center of mass of the reference marginal distribution.
On the opposite, the possibility to reach the maximum disorder of
the system strictly is affected by the dependence structure between
the in- and out-degree; such an aspect is captured through suitable
copulas.

Therefore, the present paper adds new perspectives to some
specific aspects of the existing literature. First, portfolio owners
are not considered as external to the market, but they are part of
the market. This implies the introduction of the concepts of
integration and diversification;  such an approach creates a bridge
between the literature on companies performances and the one on
companies interactions, where the embedding of a company in a
network is a key factor. Second, we base our analysis on data
available both in literature and on the case study for exploring the
configurations that lead to max/min entropy when both integration
and diversification are considered. Concentration is here intended
as the maximal correlation among diversification and integration. It differs from
the well known assortativity on networks due to the way of
measurement:
the assortativity is the correlation among diversification and
integration measured from raw data \cite{Newman}. Differently, concentration is
calculated through the entropy and under the hypotheses of different
correlation structures, expressed through copulas.

\section*{Acknowledgments} 

The Authors thank Prof. Anna Maria D'Arcangelis for
for providing data and  fruitful  discussions



\newpage 
\section*{Appendices}

\section*{A. Maximum of a product and minimum of the Shannon entropy}

{\bf Theorem A1} 
(a) Given two vectors with non negative components $p=(p_1, p_2, \cdots, p_n)$,   $q=(q_1, q_2, \cdots, q_n)$, then the minimum of the scalar product under the permutation of the components of one of the vectors is achieved for $q^\star=(q^\star_1, q^\star_2, \cdots, q^\star_n)$ i.e.: $min_{\pi \in \Pi_n} \sum_{k=1} ^n p_k q_{\pi_k}=\sum_{k=1} ^n p_k q^\star_{k}$, with reverted ranked components, i.e. $p_i\ge p_j$ and $q^\star_i \le q^\star_j$, for each $i<j$.
\newline(b)
Given two vectors with non negative components $p=(p_1, p_2, \cdots, p_n)$,   $q=(q_1, q_2, \cdots, q_n)$, then the maximum of the scalar product under the permutation of the components of one of the vectors is obtained for $q^\star=(q^\star_1, q^\star_2, \cdots, q^\star_n)$ i.e.: $max_{\pi \in \Pi_n} \sum_{k=1} ^n p_k q_{\pi_k}=\sum_{k=1} ^n p_k q^\star_{k}$, with components ranked in the same order, i.e. $p_i\le p_j$ and $q^\star_i \ge q^\star_j$, for each $i<j$.
\newline
\emph{Proof.} We report only the proof of (a), since the proof of (b) is analogous. \newline (a)It holds
$ \sum_{k=1} ^n p_k q_{\pi_k}=\sum_{k=1,k\neq i,j} ^n p_k q_{\pi_k}+p_iq_i+p_jq_j\le \sum_{k=1,k\neq i,j} ^n p_k q_{\pi_k}+p_iq_j+p_iq_j $.
In fact, $p_i q_i+p_jq_j \le p_i q_j+p_j q_i$ is equivalent to writing
$p_i (q_i-q_j)-p_j(q_i-q_j) \le 0$, that happens when $(p_i-p_j)(q_i-q_j) \le 0$, that is verified  if, anytime  $p_i \ge p_j$, then $q_i \le q_j$.

{\bf Remark A2} Results of Theorem A1 hold under the same hypothesis
and for monotonic transformations of $p$ or $q$. In particular, this
is true in case of logarithmic transformation. Now, entropy can be
seen as the inner product of two vectors: one containing the
probability, and the other its logarithm. Thus, the ranking of the
two vectors is always the same, and Theorem A1 guarantees that
entropy is maximal when the distributions are as flat as possible,
and minimal when the mass is concentrated as most as possible on
some units - attaining the true maximum for the Dirac's Delta
function.

\section*{B. Steps 2-5 of case 3}

\subsection*{Step 2:  power law for $k_{in}$, and raw data for $k_{out}$}

\begin{itemize}
    \item Non parametric copulas.
The situation is quite similar to Figure
\ref{fig:NonParametricKinFisso}. The product copula and the Upper
Frechet are quite close each to the other. The same comments as
for Figure \ref{fig:NonParametricKinFisso} hold. The functions are
decreasing as $k$ increases. There are local maxima:
    in $k=0.5$ $H=1.99 $ (Product), in $k=2$ $H=0.93 $ (Lower Frechet),
    in $k=1.3$ $H=1.86 $ (Upper Frechet).
We remark that there are many more small fluctuations, that lead to
local minima for the Upper Frechet - although the values of the
entropy there is much higher than the value on the tail. In the
lower Frechet we remark that the local minima have a different
location:    for $k=1$ $H=0.59 $ and or $k=0.5$ $H= 0.31$. There is
also a local maximum in $k=0.9$ $H=0.62 $

    \begin{figure}
        \centering
        \includegraphics[width=1\textwidth]{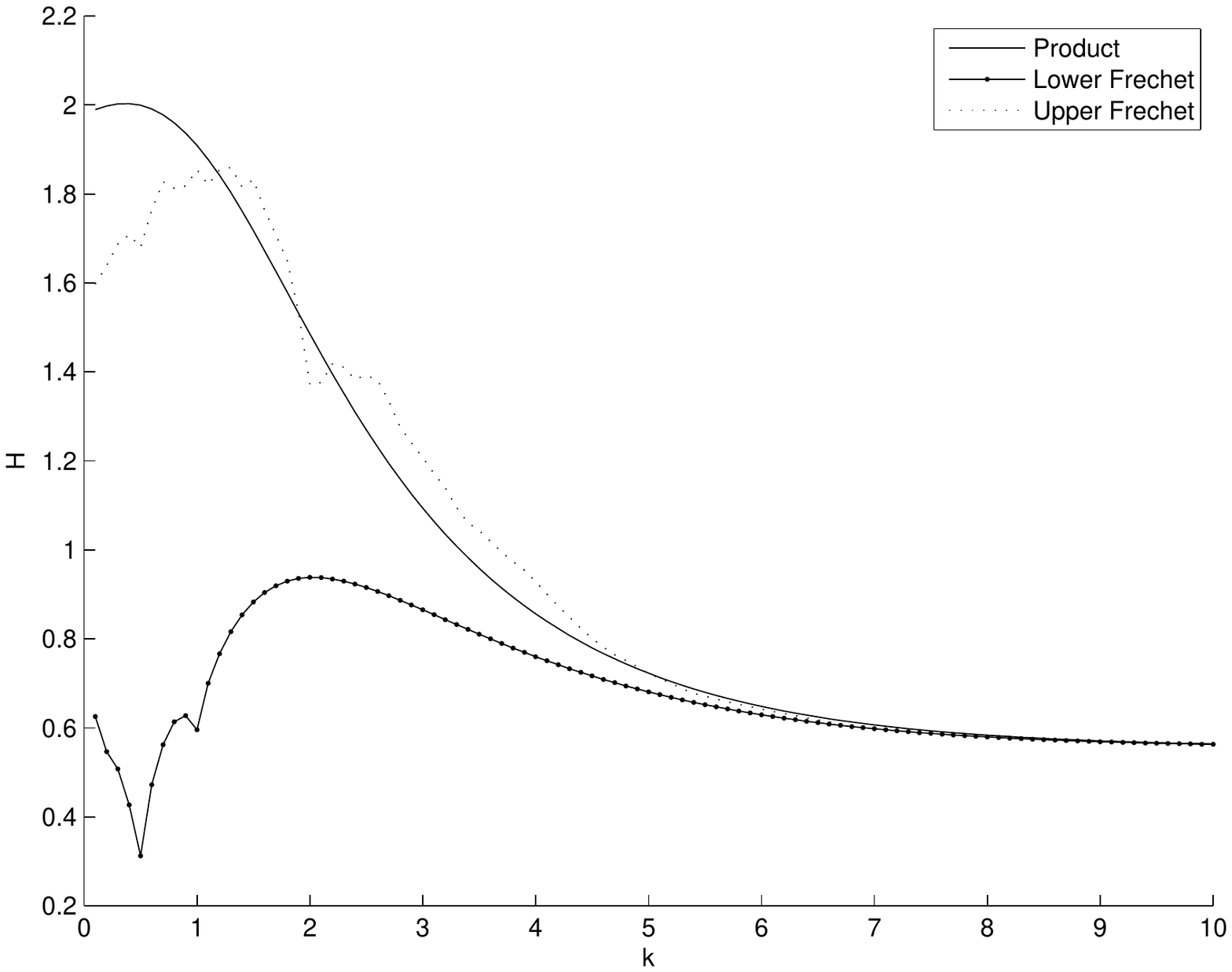}
\caption{The figure shows three cases for the entropy (y-axis) as a
function of $\theta$ (x-axis). The marginals are: the power law for
$k_{in}$ and from the case study for $k_{out}$. The situation is
quite similar to the one in Figure \ref{fig:NonParametricKinFisso},
but there are many more local fluctuations in the Upper Frechet
copula.} \label{fig:NonParamKinPL}
    \end{figure}

\item Gumbel Archimedean copula. Figure \ref{fig:KoutFissoKinPLGumbel} shows the case.  The same comments as for Figure 6 hold.
   
    \begin{figure}
        \centering
        \includegraphics[width=1\textwidth]{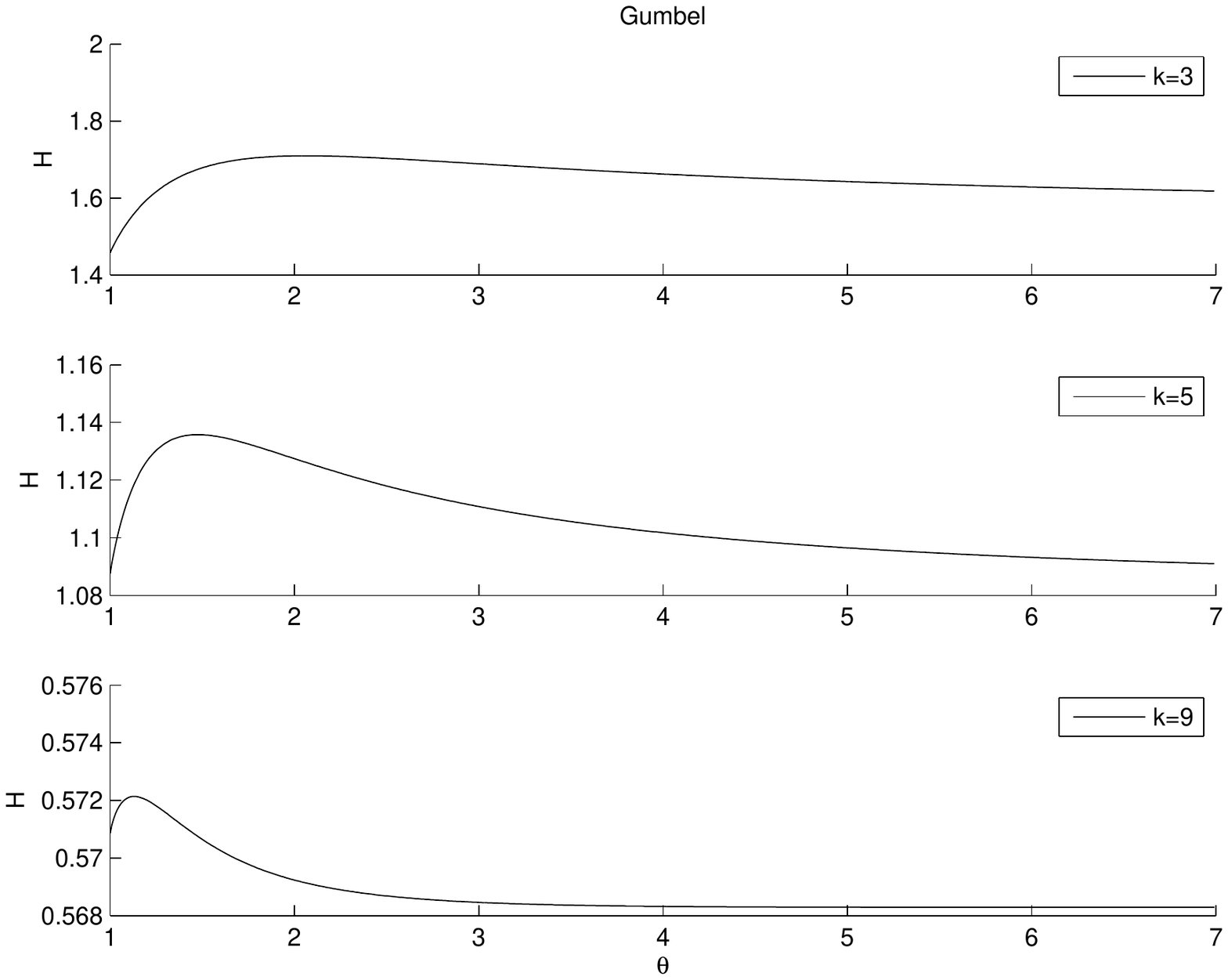}
\caption{Plot of the dependence of the entropy function $H$ on the
parameter $\theta$. The same comments as for Figure
\ref{fig:Prova2GumbelM0} hold.} \label{fig:KoutFissoKinPLGumbel}
    \end{figure}

\item Frank Archimedean copula.  Figure \ref{fig:KoutFissoKinPLFrank} shows the case.
The same comments as for $k_{in}$ from the empirical data and
$k_{out}$ power law hold (Figure \ref{fig:Prova2Frank}.).
   
    \begin{figure}
        \centering
        \includegraphics[width=1\textwidth]{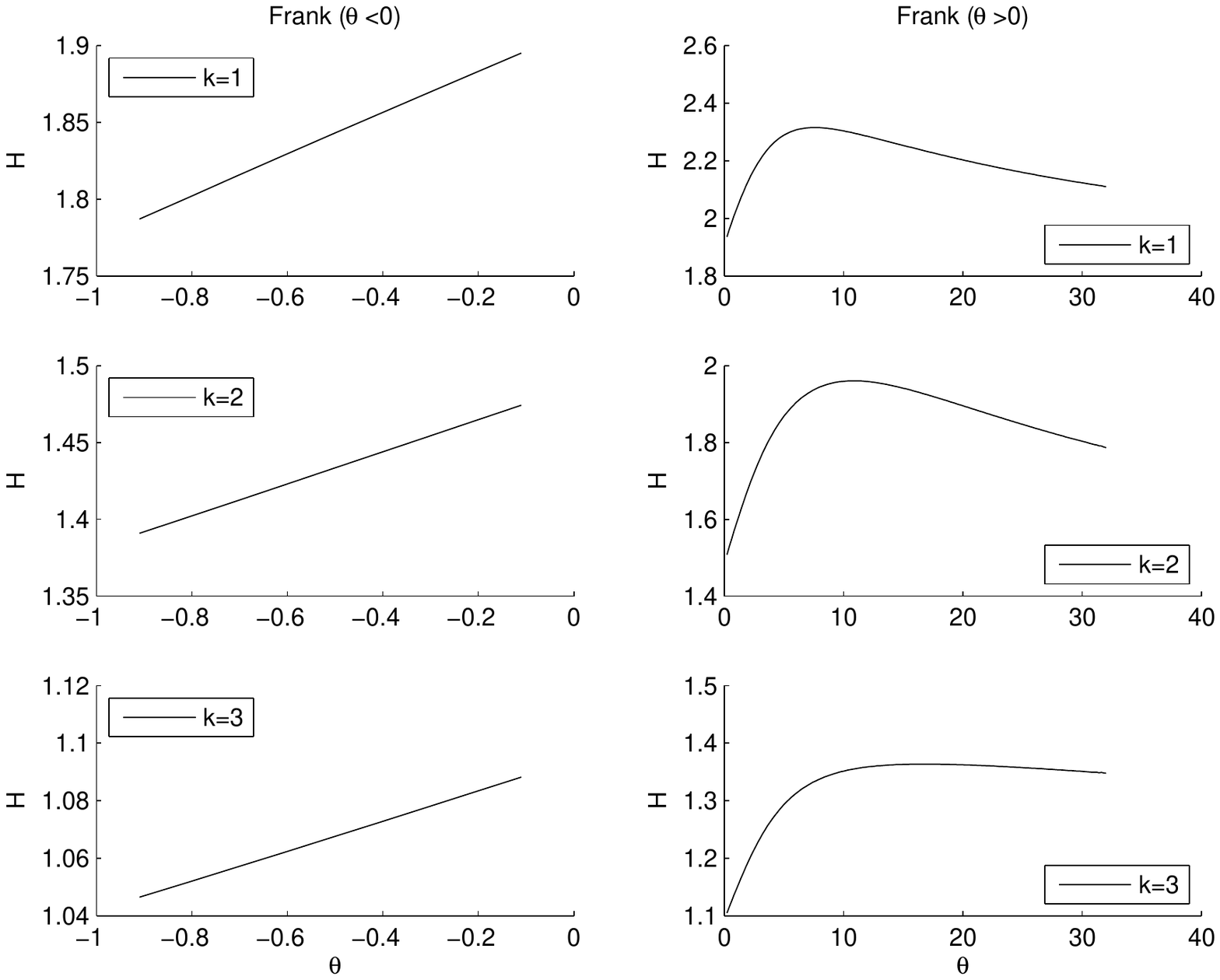}
\caption{Plot of the dependence of the entropy function $H$ on the
power law exponent $k$. The same comments as for Figure
\ref{fig:Prova2Frank} hold.} \label{fig:KoutFissoKinPLFrank}
    \end{figure}

\item Clayton Archimedean copula. Figure \ref{fig:KoutFissoKinPLClayton} shows the case.
The same comments as for $k_{in}$ from the empirical data and
$k_{out}$ power law hold (Figure \ref{fig:Prova2Clayton}).
  
    \begin{figure}
        \centering
        \includegraphics[width=1\textwidth]{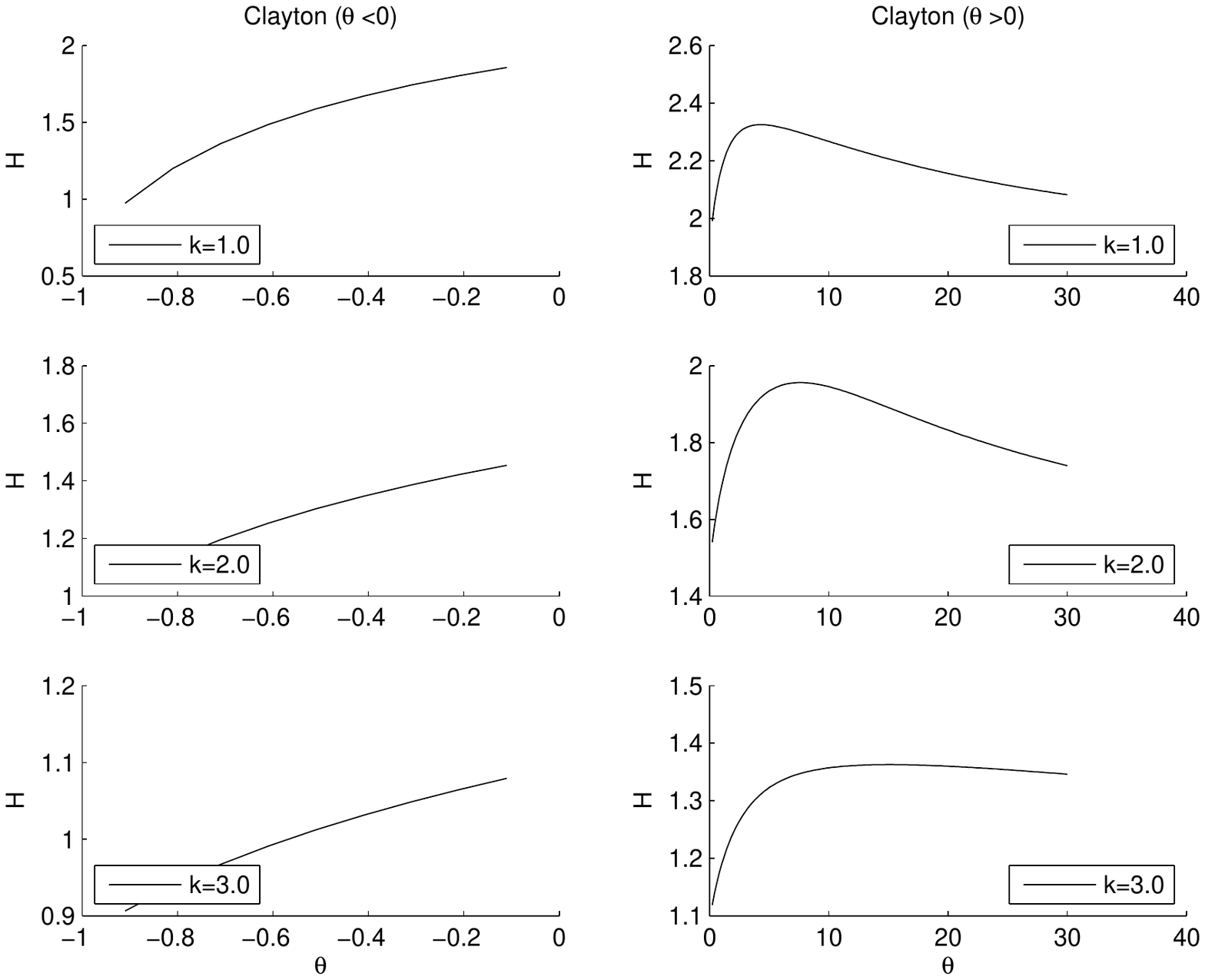}
\caption{Plot of the dependence of the entropy function $H$ on the
power law exponent $k$. The same comments as for Figure
\ref{fig:Prova2Clayton} hold.} \label{fig:KoutFissoKinPLClayton}
    \end{figure}

\end{itemize}

\subsection*{Step 3: exponential law for $k_{in}$,
 raw data for $k_{out}$}

\begin{itemize}
\item Non parametric copulas.
The situation is quite similar to Figure
\ref{fig:NonParametricKinFisso}. The product copula and the Upper
Frechet are quite close  to each other. The same comments as
for Figures \ref{fig:NonParametricKinFisso} and
\ref{fig:NonParamKinPL} hold. The functions are decreasing as $k$
increases. Figure \ref{fig:KoutFissoKinExpNonParam} shows the
results. There are local maxima:

    in $k=0.11$ $H=2 $ (Product), in $k=1.06$ $H=0.98 $ (lower Frechet),
    in $k=0.46$ $H=1.85 $ (upper Frechet).
We remark that there are many more small fluctuations, that lead to
local minima for the upper Frechet - although the values of the
entropy there is much higher than the value on the tail. Compared to
Fig. 9, the local minimum in the upper Frechet at $k=1.06$, $H=1.31$
is much deeper, and could be considered a true local minimum. In the
lower Frechet case,
we remark that the local minima have a different
location: for $k=0.41$ $H=0.63$ and or $k=0.16$ $H= 0.33$.

There is also a local maximum in $k=0.31$ $H=0.66 $

    \begin{figure}
        \centering
        \includegraphics[width=1\textwidth]{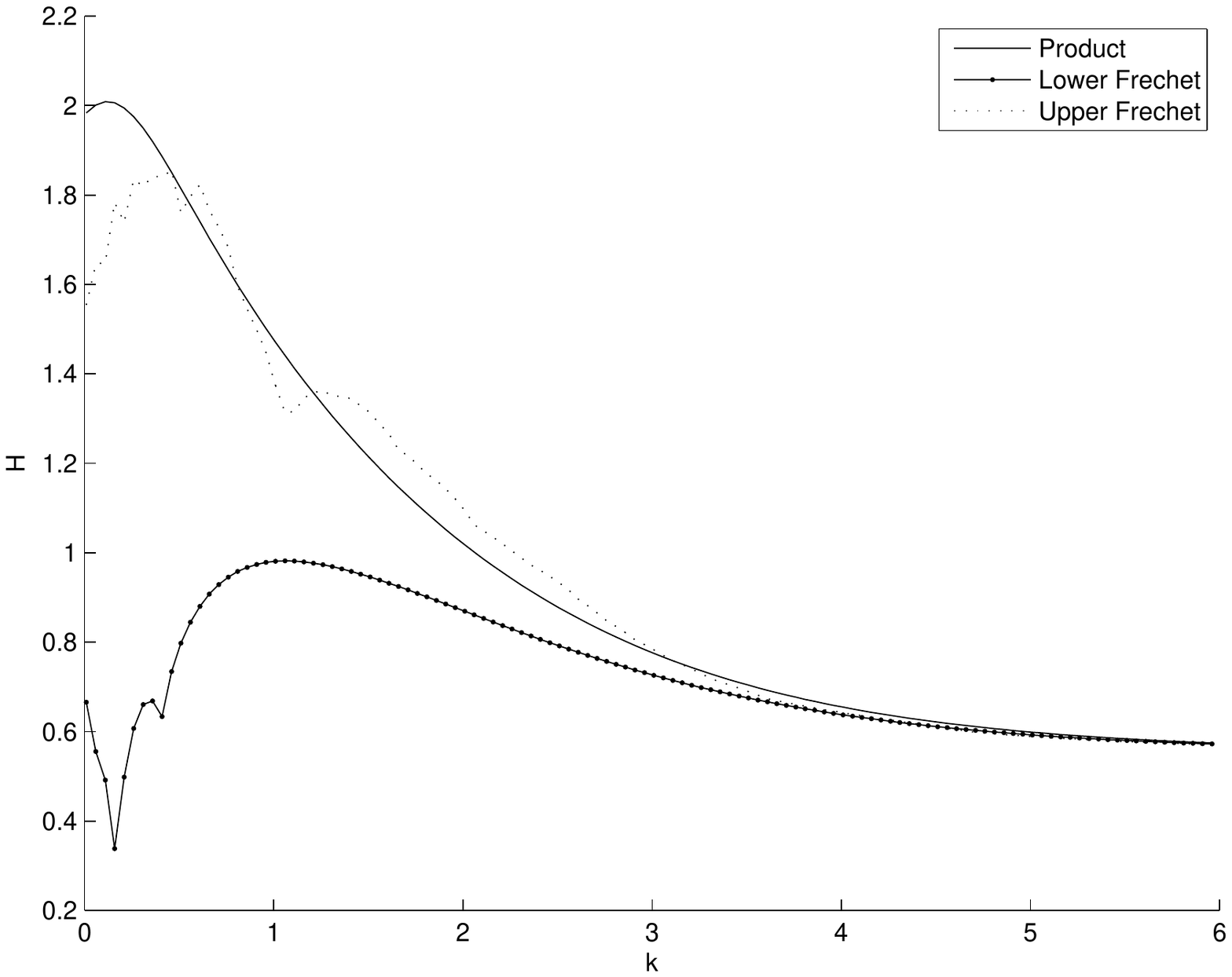}
\caption{The figure shows three cases for the entropy (y-axis) as a
function of $\theta$ (x-axis). The marginals are: the power law for
$k_{in}$ and from the case study for $k_{out}$. The situation is
quite similar to the one in Figure \ref{fig:NonParamKinPL}, but the
local fluctuations in the upper Frechet copula are deeper. }
\label{fig:KoutFissoKinExpNonParam}
    \end{figure}

    \item Gumbel Archimedean copula. Figure \ref{fig:KoutFissoKinExpGumbel} shows the case.  The same comments as for Figures  6 and  10 hold.
    
    \begin{figure}
        \centering
        \includegraphics[width=1\textwidth]{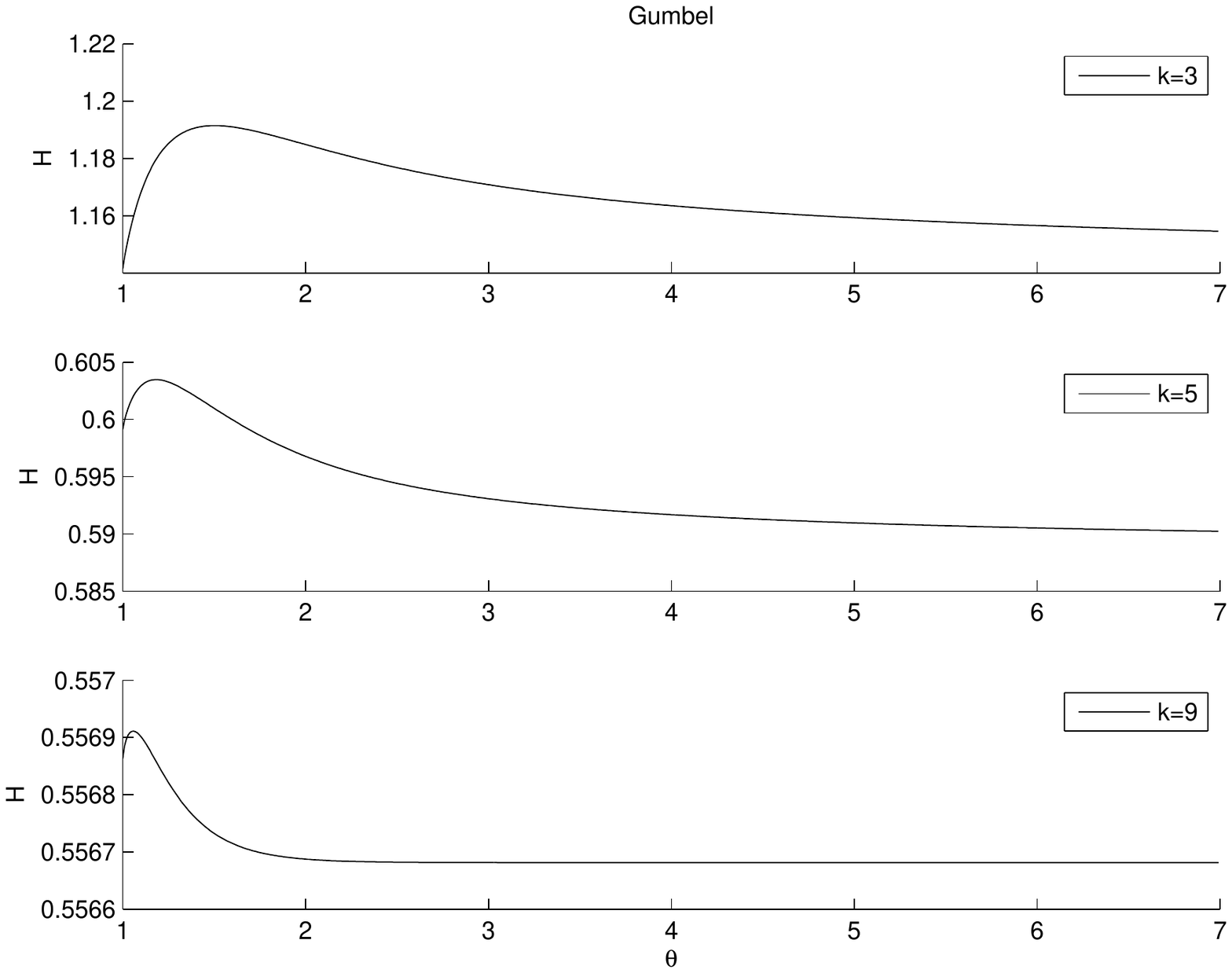}
\caption{Plot of the dependence of the entropy function $H$ on the
parameter $\theta$. The same comments as for Figure
\ref{fig:KoutFissoKinPLGumbel} hold.}
\label{fig:KoutFissoKinExpGumbel}
    \end{figure}

\item Frank Archimedean copula.  Figure \ref{fig:KoutFissoKinExpFrank} shows the case.
The same comments as for Figures \ref{fig:Prova2Frank} and
\ref{fig:KoutFissoKinPLFrank}
 hold.
 
    \begin{figure}
        \centering
        \includegraphics[width=1\textwidth]{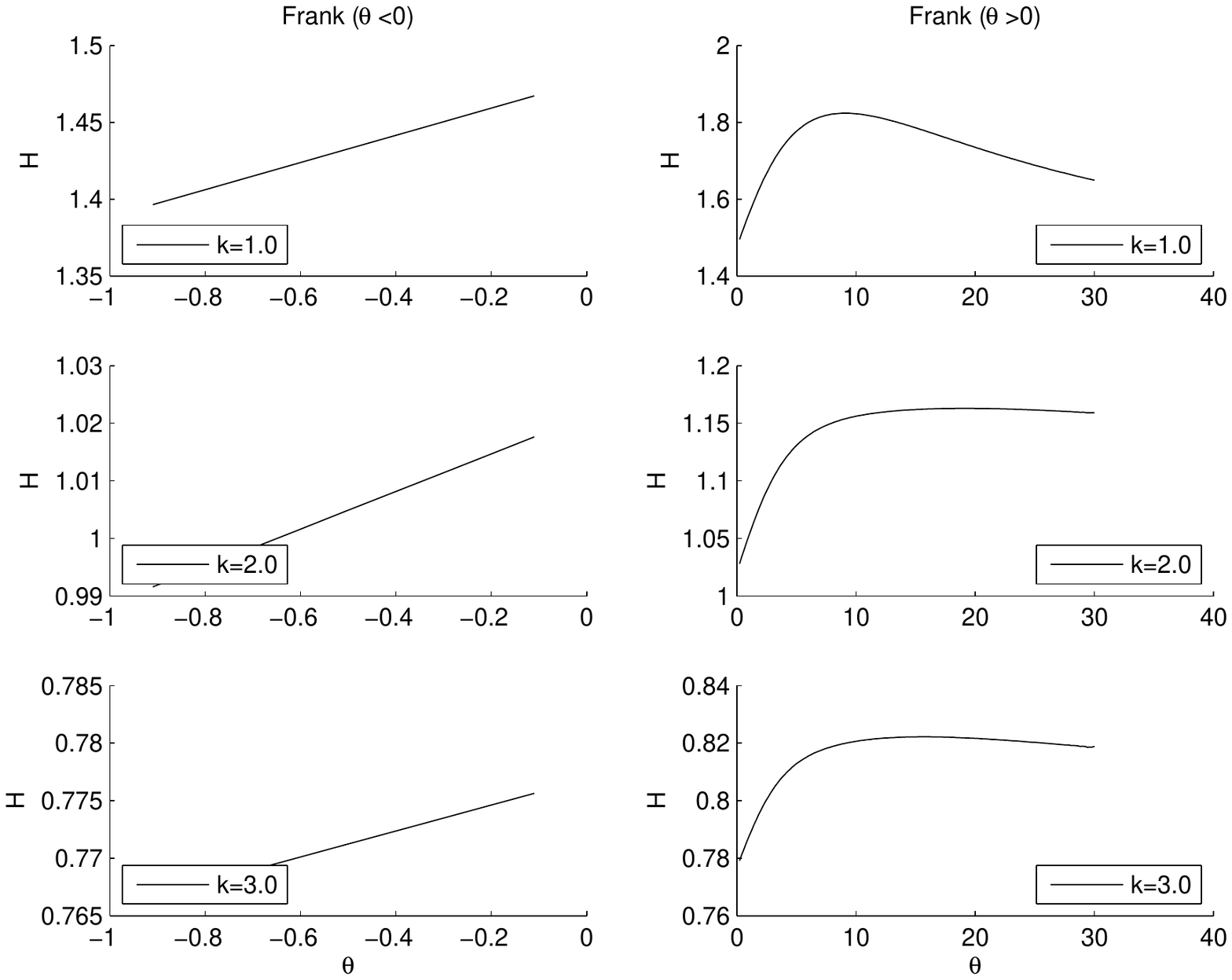}
\caption{Plot of the dependence of the entropy function $H$ on the
parameter $\theta$.  The same comments as for Figures
\ref{fig:Prova2Frank} and \ref{fig:KoutFissoKinPLFrank} hold.}
\label{fig:KoutFissoKinExpFrank}
    \end{figure}

\item Clayton Archimedean copula. Figure \ref{fig:KoutFissoKinExpClayton} shows the case.
The same comments as for Figures \ref{fig:Prova2Frank} and
\ref{fig:KoutFissoKinPLClayton} hold.
    \begin{figure}
        \centering
        \includegraphics[width=1\textwidth]{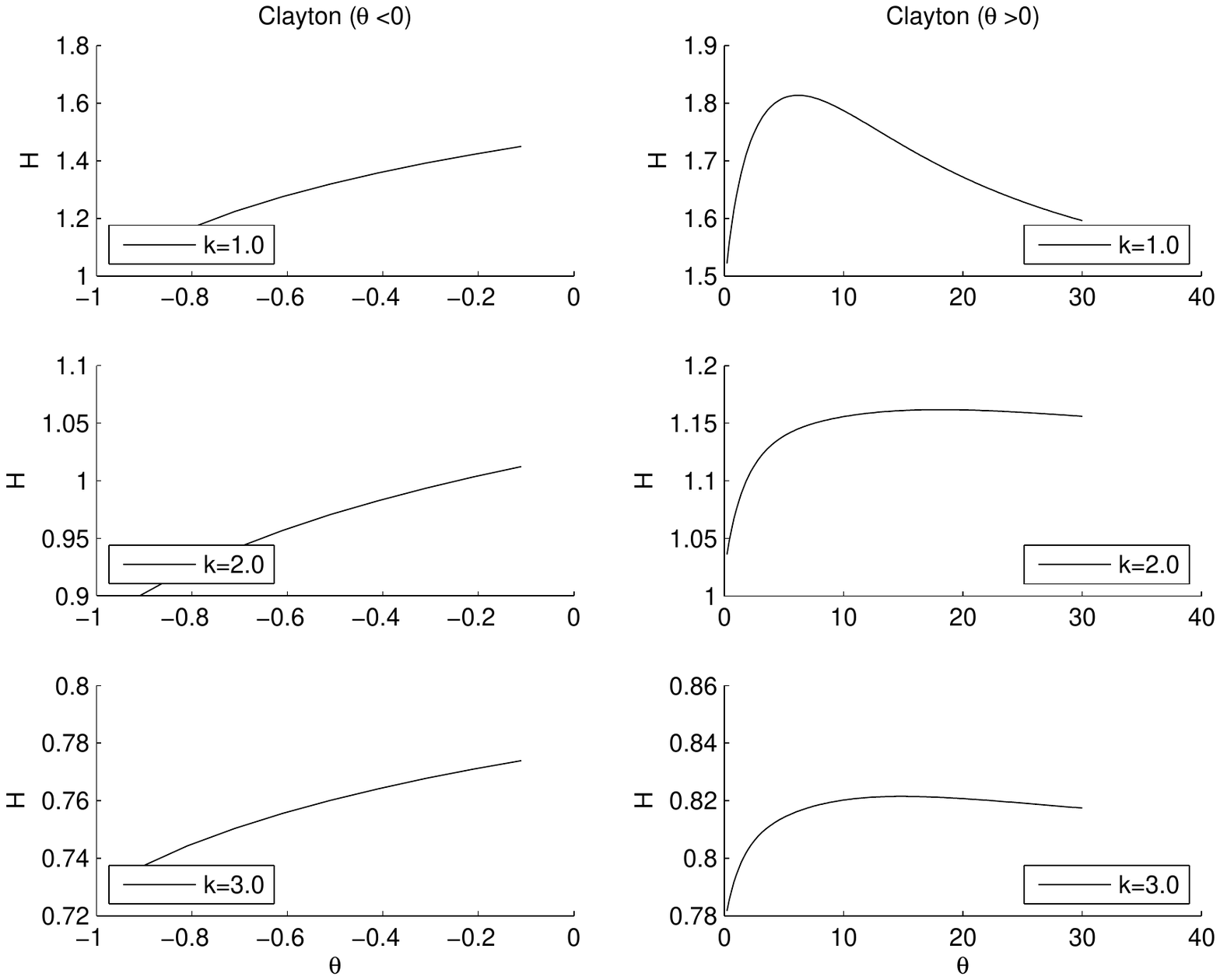}
\caption{Plot of the dependence of the entropy function $H$ on the
parameter $\theta$. The same comments as for Figures
\ref{fig:Prova2Frank} and \ref{fig:KoutFissoKinPLClayton} hold.}
\label{fig:KoutFissoKinExpClayton}
    \end{figure}

\end{itemize}

\subsection*{Steps 4 and 5:
either power law or exponential law for $k_{in}$, and power law for
$k_{out}$} On the parametric copulas, in view of the numerical
results already obtained, of the Theorem A1, and due to Remark A2 in
the Appendix, in   cases of either power law or exponential law for
$k_{in}$, while $k_{out}$ remains described by a power law, we
conclude that the entropy diminishes as the parameters for the power
law(s) or the exponential go to infinity. There will be local maxima
that will go either to the left or to the right border of the range
of $\theta$ as the power law/exponential parameters increase.

\end{document}